# Horizon Scans can be accelerated using novel information retrieval and artificial intelligence tools

Lena Schmidt[1], Oshin Sharma[1], Chris Marshall[1], Sonia Garcia Gonzalez Moral[1]

Affiliation: NIHR Innovation Observatory, Population Health Sciences Institute, Newcastle University, Newcastle upon Tyne, UK

**Introduction:** Horizon scanning in healthcare assesses early signals of innovation, crucial for timely adoption. Current horizon scanning faces challenges in efficient information retrieval and analysis, especially in grey literature, presenting a need for innovative tools.

**Methodology:** This article introduces SCANAR and AIDOC, open-source Python-based tools designed to improve horizon scanning. SCANAR automates the retrieval and processing of news articles, using de-duplication and unsupervised relevancy ranking. AIDOC aids filtration by leveraging large language models and text embeddings for calculation of semantic similarity, to reorder data based on relevancy.

**Results:** Twelve internal datasets from horizon scans and four external benchmarking datasets were used. SCANAR improved retrieval efficiency by automating processes previously dependent on manual labour. AIDOC displayed work-saving potential, achieving around 65% theoretic reduction in manual review efforts at 95% recall or 20% reduction when adding CMH, an existing open-source algorithm to make data-driven decisions on stopping screening early. Comparative analysis with benchmarking data showed AIDOC's performance was similar to existing systematic review automation tools, though performance varied depending on dataset characteristics.

**Conclusion:** The validation indicates that SCANAR and AIDOC show potential to enhance horizon scanning efficiency by streamlining data retrieval and prioritisation. These tools may alleviate methodological limitations and allow broader, swifter horizon scans. Further studies are suggested to optimize these models and to design new workflows and validation processes that integrate large language models.

# Introduction

## Horizon scanning background

Horizon scanning is an analytical foresight technique that is used across diverse sectors. Its main purpose is to systematically examine information sources to identify early signs of important developments (or weak signals) and better anticipate possible futures (Cuhls, 2020; Hines et al., 2019). In healthcare, horizon scanning methods are particularly useful to support the detection and prioritisation of innovative medicines and other medical technologies including devices, diagnostics and digital health applications and systems (MedTech) ahead of these technologies being placed on the market (Cuhls, 2020; Garcia Gonzalez-Moral et al., 2023; Hines et al., 2019). This anticipatory intelligence allows interest holders and decision-makers to prepare for the adoption of such technologies. This process becomes crucial for example in the case of disruptive technologies such as Artificial Intelligence (AI) that have the potential to change entire healthcare pathways (Reddy, 2024). Given the strategic value of horizon scanning in anticipatory policy and decision making in healthcare, timeliness is of upmost importance in this context as delay in the identification of innovative healthcare technology may pose a risk to patients' health and incur in potential efficiency losses or increased healthcare costs.

Horizon scanning relies on systematic approaches for weak signal identification and follows a well set-out process, although there may be room for pragmatic adaptation of the general workflow for urgent or broad-scoped scans (see Figure 1 for a description of the general steps). Horizon scanning outputs, for example those from the NIHR Innovation Observatory (IO), are reported and disseminated to interest holders. This includes not only research commissioners but is also inclusive of those who will potentially benefit from early intelligence - such as patients, professional organisations, or the research community.

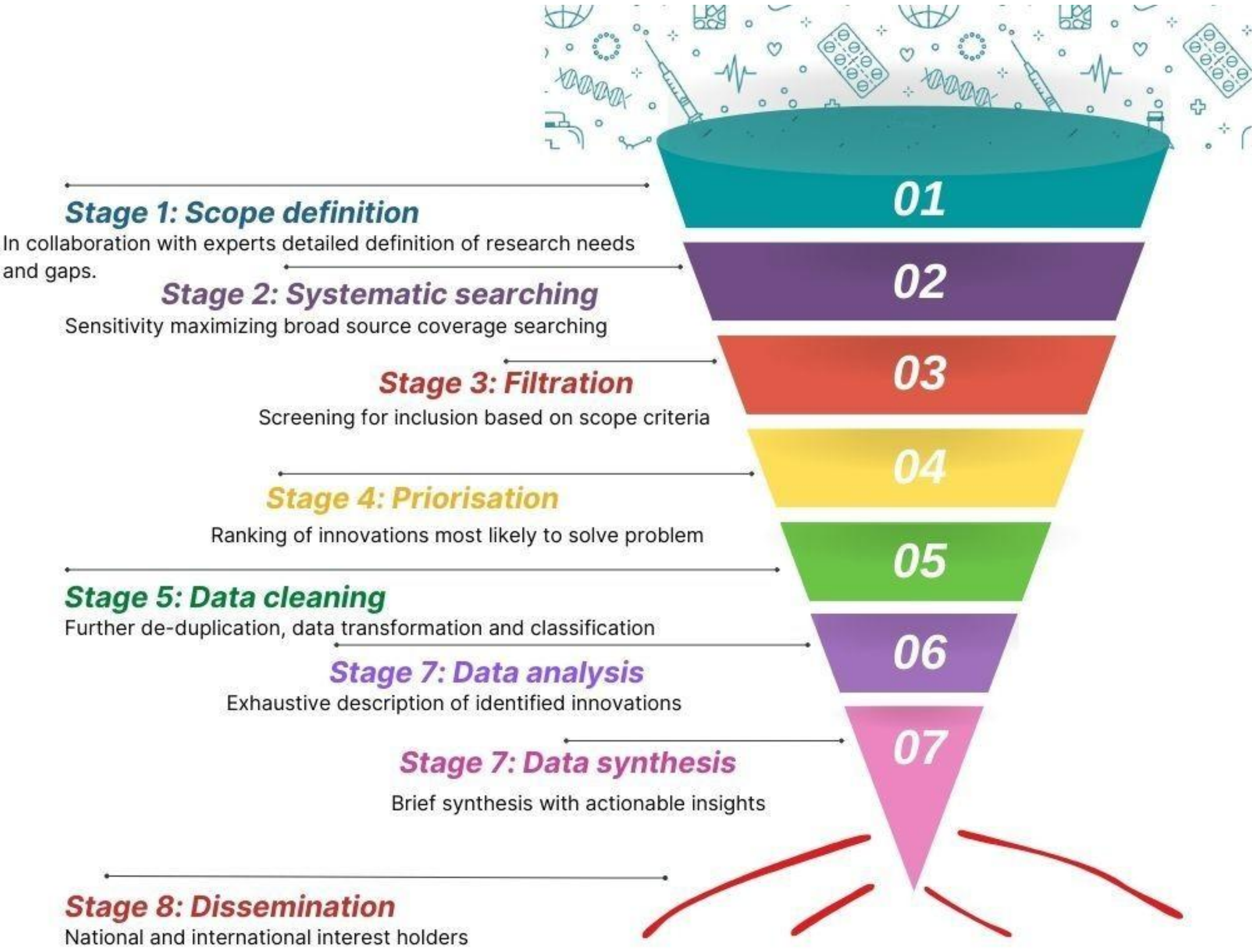

*Figure 1: Horizon scanning stages. The stages were adapted and extended from Simpson et al. (2014)*

Unlike in systematic approaches to evidence synthesis that start by defining the topic in scope, horizon scans are often searching beyond the known domain. It includes 'looking into the future' as comprehensively as can be by maximizing the sensitivity of the searches and the number of sources searched (van Rij, 2010).

For this reason, horizon scanning presents information retrieval challenges. Firstly, some sources, such as news sites, can be quite resource-intense and laborious for information retrieval tasks. This is because they do not present advanced search interfaces or download features. They may present filtering or sorting options to display the list of results. This lack of advanced functions makes the search process very time consuming because many different search terms need to be added independently due to the lack of Boolean operators.

Horizon scans of news articles, for example, are affected by this problem. Their search results may need to be 'triaged' for inclusion directly on screen, separately for each search query. This leads to mixing up the horizon scanning stages 2+3 (Systematic searching and Filtration), which is not ideal in terms of reproducibility and transparency. Lengthy processes and duplication are side-effects of this workflow, since the same news item might be retrieved by the multiple separate searches. The main cause of this problem is the lack of basic download or export functions. This prevents information specialists from downloading all results in spreadsheets or other formats that are compatible with digital tools that help with managing the workflow of further assessment and analysis. These manual copy and paste operations negatively impact transparency or systematicity for future reproducibility. Secondly, relying on manual processes for information retrieval denies the ability to collect performance metrics (e.g. precision and sensitivity of the search) because the number of search results or number retrieved for each search is not recorded. This, in turn, impacts our ability to estimate the time needed to perform filtration and data-extraction tasks that come later in the horizon scanning process. For the remainder of this manuscript we will use the term ‘Filtration’ to refer to the process of sifting or screening relevant information for inclusion in a horizon scan.

The use of machine learning in evidence synthesis, especially for systematic and living reviews, is a growing field of research with new tools constantly emerging (Johnson et al., 2022). However, automation has an even more obvious potential in horizonscanning due to the rapid nature of methods and the use of unstructured soft intelligence from sources such as news, patents, social media, websites, and others. This soft, or ‘grey literature’ contains the emerging signals that horizon scans aim to discover, but the quantity of documents and texts from these sources are very large. The application of (semi)automated screening and data extraction in systematic reviews are active topics of research (Kusa, Zuccon, et al., 2023; Schmidt et al., 2024), and potentially game-changing technologies such as large language models (LLMs) are being tested for their suitability to expedite literature review and evidence synthesis across multiple steps of the workflow (Lieberum et al., 2025). To the best of our knowledge, however, there are no dedicated and currently available tools for accelerating or automating horizon scanning information retrieval and filtration, or data extraction processes.

Within the Innovation Observatory we are actively working on filling this gap. We are currently developing and validating a set of modular open-source tools to form a horizon scanning toolkit that supports the production of horizon scans across all workflow steps outlined in Figure 1.

This article is of interest to national and international interest holders in the healthcare sector, as a standardization of methods and description of evaluation methodology for horizon scanning automation; adapted from well-established methods used in the automation of systematic reviews.

## Aims of this paper

This paper aims to present solutions to the current challenges encountered at searching and filtration stages in horizon scanning (stages 2 and 3 in Figure 1). We introduce two custom-built, open-source, semi-automated and flexible horizon scanning tools developed for the information retrieval of news and support of general reference filtration. These tools were developed to support the Innovation Observatory horizon scanning centre in their routine horizon scanning tasks. In this article, we describe their validation and usage across multiple scanning projects in varying areas of health and care research, using corpora with both peer-reviewed and grey-literature. Furthermore, we conduct an evaluation and comparison to external results from the published literature, applying our methods to data from the related field of systematic review screening automation.

# Methods

## SCANAR: A Tool for Bulk Retrieval of News Articles and the Identification of Weak signals

This section describes methods for the SCANAR (Search Companion for Advanced

News Article Retrieval) tool. It is an open-source, Python-based end-user tool that automates interactions with Google News RSS feed search for the purpose of information retrieval in news-based horizon scanning research. A visual representation of the information flow through this tool is shown in Figure 2.

GoogleNewsFeed[1] and googlenewsdecoder[2] are the two main open-source Python libraries that are used to facilitate searches and information retrieval via GoogleNews.[3] SCANAR user-interface was programmed using the Streamlit library[4] and its code is available via GitHub (*see Data Availability Statement*).

SCANAR's core functionalities extend beyond simple information retrieval from Google News, by is to automating the retrieval of multiple queries related to one research question in bulk, de-duplication of articles, full text scraping, self-supervised relevancy-ranking, and data conversion and export for downstream processing such as manual review or usage of further automation tools. Its main scientific contribution is to enable systematic news article searches and to improve their reproducibility and timeliness.

**Retrieval:** Users provide a text file with multiple search queries (one query per line) via the tool's user-interface. Parameters such as time frame of the search and number of articles to retrieve per query can be selected. Each search query is then posted separately to the Google News RSS API (Application Programming Interface), which returns up to 100 results. For each query, the numbers of results are stored as part of a search documentation. Article metadata as well as the name of the query for which an article was retrieved and page number are stored as part of the results. By 'page number' we refer to the virtual paging commonly used in search engines, where for example the top 10 results are displayed on page 1 of the results; and results 11-20 are on page 2. The tool waits three seconds between queries to avoid being blocked by Google News for excessive API requests.

**De-duplication:** Since search queries are related to the same research question, it is to be expected that multiple similarly worded queries may return the same relevant article (i.e. a duplicate). De-duplication runs in parallel with the information retrieval process and later plays key part in informing self-supervised ranking within the tool. For each article retrieved from the RSS feed, a concatenation of title and URL are used to match new articles with articles retrieved via previous queries in a session.

[1] https://github.com/LLukas22/Google-News-Feed (last accessed 19/03/2026)
[2] https://github.com/SSujitX/google-news-url-decoder (last accessed 19/03/2026)
[3] https://news.google.com/home?hl=en-GB&gl=GB&ceid=GB:en (last accessed 19/03/2026)
[4] https://streamlit.io/ (last accessed 19/03/2026)

Each article that is identified as duplicate is not added as new entry, but the original article's entry in the search results is updated by incrementing a counter of how many times it was identified as duplicate. Paging is the concept of returning search results in batches, for example when using search engines, the first relevant websites appear on page 1 and subsequent results follow on the next pages. In our tool, if a duplicate news article has a better (i.e. lower) page number, the original article's page number is updated to reflect the best page number across all search queries. Additionally, the search query that produced the duplicate is added to a list of search queries associated with the article.

**Full-text scraping:** If the user selects the full-text scraping option, the tool attempts to resolve the URL provided in the RSS feed and to scrape article full text from a website. Between each call to resolve a URL, the tool waits 1.5 seconds due to API limits. It only retrieves full texts for the first occurrence of each article and truncates the length of each article to 30,000 characters maximum to avoid downstream issues with Excel's maximum cell size limit after data export.

**Self-supervised relevancy-ranking:** For each unique article the minimum page number and the number of duplicates across all searches are collected. The minimum page number reflects Google's 'black box' assessment of how relevant an article is with respect to a search query. The number of duplicates reflects how well the article relates to the different searches proposed by the information specialist, and an article that was retrieved more than once is considered more relevant because it matched several different facets of the research question expressed via different searches. Articles are then re-ordered primarily on page number, and secondary re-ordering is carried out to ensure that articles with most duplicates for each minimum page number are displayed on top.

**Data export:** Data are exported in three different files. A CSV (Comma-Separated Values) file with a search documentation containing each search query, the number of results, the number of retrieved results, and the number of new unique results contributed after de-duplication for each search is provided. Secondly, another CSV file with the ranked articles, metadata, scraped full-texts if applicable, and ranking variables is provided. Thirdly, each article is converted to a news-type RIS entry and appended to a text file that can be used for import into a reference management, screening, or filtration software.

**Deployment:** For in-house use, the finished application was deployed on an Amazon server (AWS EC2 instance). Alternatively, it could have been deployed directly on the Streamlit community cloud which, at the time of writing this article, is a free service.[4]

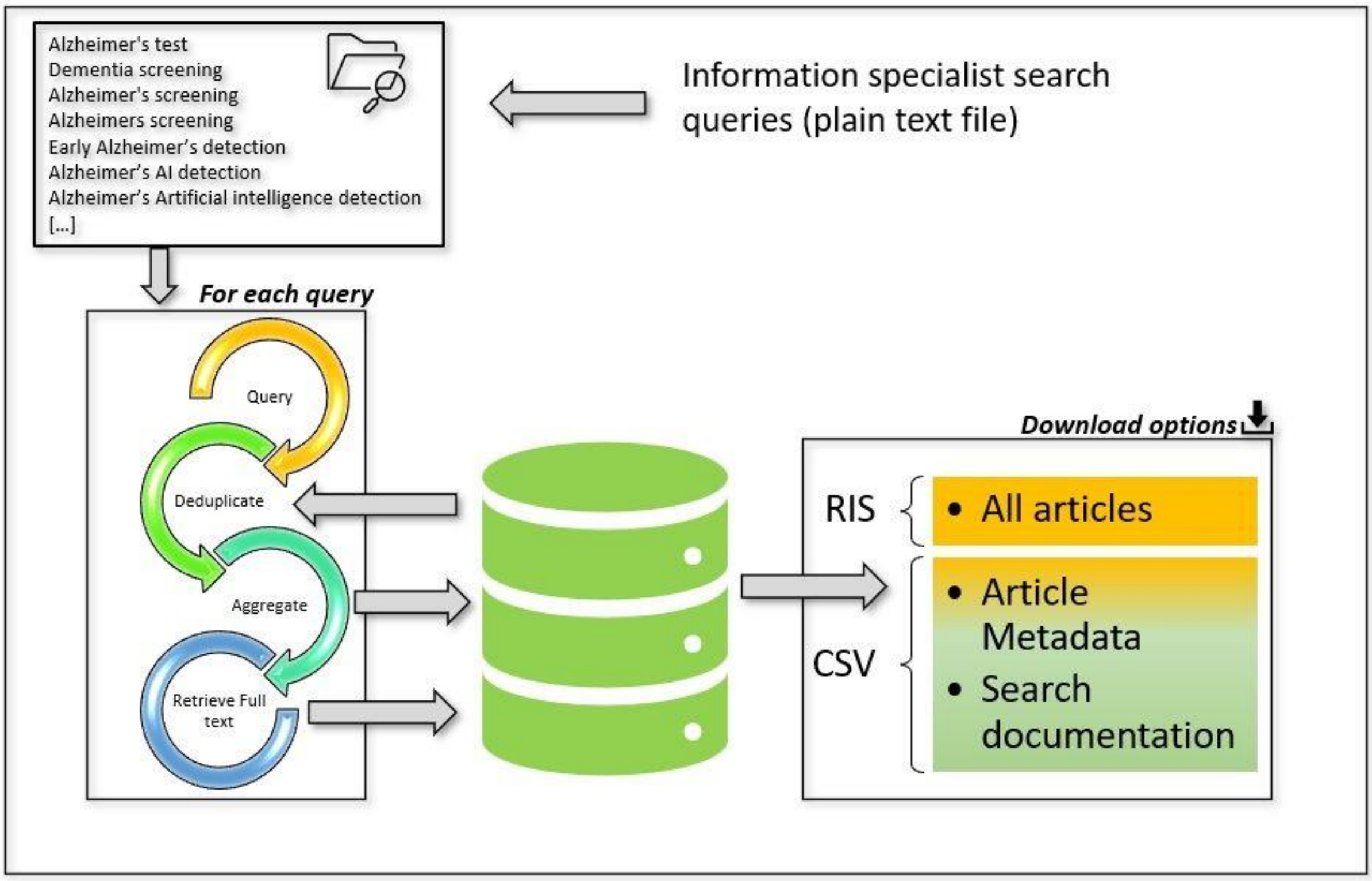

*Figure 2: SCANAR flowchart showing the information retrieval process within the tool. Firstly, a text file with one query on each line is uploaded to the tool. For each query, results are retrieved, de-duplicated, aggregated, and optionally full text is retrieved. The data is stored in an internal dataframe and can be exported as RIS, CSV. Additionally, SCANAR creates an automated search documentation that shows how many results were retrieved, and how many duplicates were contained in each query.*

## AIDOC: AI-assisted Weak Signal Identification Tool to Support Efficient Horizon-scanning Processes

In the following section we describe a highly flexible, open-source, Python-based end-user tool that assists horizon scanning analysts with filtration of results from heterogeneous sources that are commonly included in horizon scans, such as clinical trial registry entries, news, patents, funding calls, websites, or peer-reviewed literature. We named it AIDOC (Artificial Intelligence Document Organiser and Classifier).

The tool was created using open-source Python libraries such as Sentence Transformers (Reimers & Gurevych, 2019) and Streamlit and employs a neural network called SPECTER that was pre-trained on scientific text (Arman Cohan et al.,

[4] https://streamlit.io/cloud (last accessed 19/03/2026)

2020). SPECTER was chosen because it was further pre-trained and evaluated on tasks for document similarity, which includes predicting the relatedness of documents using a triplet loss function (A Cohan et al., 2020). It aims to represent whole documents, which aligned well with the aim of this work. While horizon scans do include peer-reviewed literature, for which SPECTER was set up, they also include grey literature in varying proportions. SPECTER, building on SCIBERT (Beltagy et al., 2019) which was itself built on scientific text, presented the most rigorously evaluated of such models and a comparable pre-trained model focussing solely on grey literature was not identified at the time. It may be noted that while not necessarily being peer-reviewed, the texts in horizon scanning corpora may still be broadly considered scientific in nature and thus related to the data used in both SCIBERT and SPECTER. With the fast-changing pace of developments in the field of natural language processing, it may be advisable for readers to check latest models and to replace them in the code provided as part of this manuscript if they wish to conduct their own experiments.[5]

The core functionality of the tool is to iteratively re-order rows within a spreadsheet based on a limited number of labels of relevancy or categories previously assigned by humans. This, in turn, enables reviewers to look at a prioritised list of records, with likely relevant evidence being shown first.  A visual representation of the information flow through this tool is shown in Figure 3: Flow of information through the AIDOC tool.

**User input:** Within the user-interface, a CSV file containing the results of information retrieval for a scan can be uploaded. In this file, each column is a meta- or user-data variable while each row contains a new record. At this point, the user is expected to have identified at least three relevant records for an overall in- or exclusion question, or at least three relevant examples for a member of a category. In the user interface, the user will be able to select the column that contains manually labelled data and is prompted to tell the tool which values to use as input for the AI; e.g. the value ‘Include’ in a column named ‘Screening decision’, called the ‘label’ hereafter. The user must then select the text on which the reference re-ordering should be based on; e.g. text in a column called ‘Patent abstract’. This text is referred to as ‘reference text’ hereafter.

**AI-based re-ordering:** The tool then uses the SPECTER neural network to transform reference text into a vectorized format (also called embedding). This allows computations to be carried out on a numeric and context-specific representation of text data (Devlin et al., 2019). All vectorized inputs are multiplied with each other to create a cosine similarity matrix, which is a well-established method for quantifying

---

[5] https://sbert.net/docs/sentence_transformer/pretrained_models.html with models to be changed in these lines of code: https://github.com/L-ENA/SiftingSimulations/blob/6606d79f6e0986f776747d293c99bb9c4e112f94/main.py#L218 and fallback: https://github.com/L-ENA/SiftingSimulations/blob/6606d79f6e0986f776747d293c99bb9c4e112f94/neural_classifier.py#L36-L37 (last accessed 19/03/2026)

semantic similarity between input and reference word embeddings (Mikolov et al., 2013) that has been extended to sentences (Reimers & Gurevych, 2019). Taking text from rows labelled as relevant by the user as reference-standard, the tool then reorders the remaining documents such that rows with the highest cosine similarity appear on top. AIDOC selects a maximum of 10 random relevant references for each reordering.

**Iterative learning:** After manually labelling the most likely data on top and identifying new relevant records, the user can then repeat the reference reordering to retrieve a new batch of likely includes. This is an iterative process, as argued by Repke et al. (2026), but is also commonly referred to as 'Active learning' in the systematic review automation literature (Ferdinands et al., 2023; Marshall & Wallace, 2019; Singh et al., 2018).

**Ensemble model using machine-learning:** Besides the basic AIDOC model that reorders references based on semantic similarity based on 10 random includes, we created an ensemble model that incorporates a machine-learning model using Term Frequency – Inverse Document Frequency (TF-IDF) vectorisation and Stochastic Gradient Descent (SGD) using log-loss, adapted from an existing Git-Hub repository (DESTINY project) containing code originally provided by the EPPI-Centre.[6] In this combined model, every fifth re-ranking is carried out by the TF-IDF/SGD model, while the other four are re-ranked using the original ranking method. This enables the combined model to focus on separate facets of the included records by concentrating on 10, while also periodically taking advantage of the entirety of all relevant records. To avoid an imbalanced training dataset, our adaptation of the SGD model is supplied with a maximum ratio of 3:1 randomly chosen negative labelled examples per each include. In other words, if the iterative learning process uncovered 15 includes then a maximum of 45 excludes are added to the training dataset of the classifier. Before each usage of the SGD, a gridsearch is carried out to determine the alpha parameter.

**LLM ensemble model:** The experimental LLM ensemble model utilises binary LLM predictions in conjunction with the standard iterative learning process outlined above, with the aim of further boosting likely relevant records to be assessed first. In practice, it requires users to formulate prompts based on a structured template, provided in Table 1. We experimented with using 'gpt-4o-mini-2024-07-18' with knowledge cut-off on Oct 01 2023, as well as 'gpt-5-mini-2025-08-07' with May 31$^{st}$ 2024 cutoff via the OpenAI API.[7] We obtain binary inclusion labels and assessment justifications using

[6] Code adapted from https://github.com/destiny-evidence/stoppingmethods/blob/main/rankings/simple.py and described here https://eppi.ioe.ac.uk/CMS/Portals/35/machine_learning_in_eppi-reviewer_v_7_web_version.pdf (last accessed 19/03/2026)

[7] https://openai.com/index/gpt-4o-mini-advancing-cost-efficient-intelligence/ (last accessed 19/03/2026)

the structure described in Table 1 and fed them into the iterative learning system as additional system inputs (besides reference text and goldstandard label). With each step in the iterative learning simulation, AIDOC then adds up its own cosine-similarity based relevancy score, which is a continuous number between 0-1, with the binary LLM result which is either 0 or 1. The resulting list therefore contains continuous scores, with the most likely-rated AIDOC references that were also prioritised by the LLM to be processed first in the simulation. All prompts were developed 'blindly', by a single person who transferred inclusion criteria into the prompt template without conducting any test runs or further refinement. Prompts used for each dataset are available in the appendix (see file managed_runs.json).

*Table 1: Suggested LLM prompt structure, broken down into relevant parts. The final prompt is always a single piece of text, as shown in our shared code repository. The example in this table shows a potential classification query provided to the LLM. For each dataset (ie. scan), prompt parts 1-3 need to be adjusted. For each piece of literature that is sent off for classification, prompt part 5 needs to be filled in.*

| Prompt Part | Content | Example |
|---|---|---|
| **1: Setting the scene** | Description of research context | "You are a researcher screening news articles for inclusion in a literature analysis." |
| **2: Summary of relevant information** | Description of project and relevant information within references | "The inclusion criteria are the following: Any article that describes a newly developed or upcoming health screening method or campaign for the early detection of diseases. Screening tests can be diagnostic; to detect cancer, dementia, HPV, or any other disease and health condition within a population. Screening methods may be offered or evaluated based on a whole population or people of selected age groups and locations. Any method, such as at-home, point of care, AIsupported, analysis of biomarkers and genes, or other methods are of interest, as long as they aim to detect diseases early." |
| **3: Exclusion criteria (optional)** | Description of irrelevant references | *escription of irrelevant evidence if le, eg.: "Exclude articles that describe established screening programmes.", articles that screen for disease XYZ.", news articles that solely describe d academic papers."]* |
| **4: Output structure** | Description of binary output format | “Answer YES if the article is relevant or unclear. Answer NO if it is not. Then reproduce the exact context from the paper that contained the information on which basis you made the decision.” |
| **5: Context** | The relevant reference text | “Here is the text of the article: [*insert text*]” |

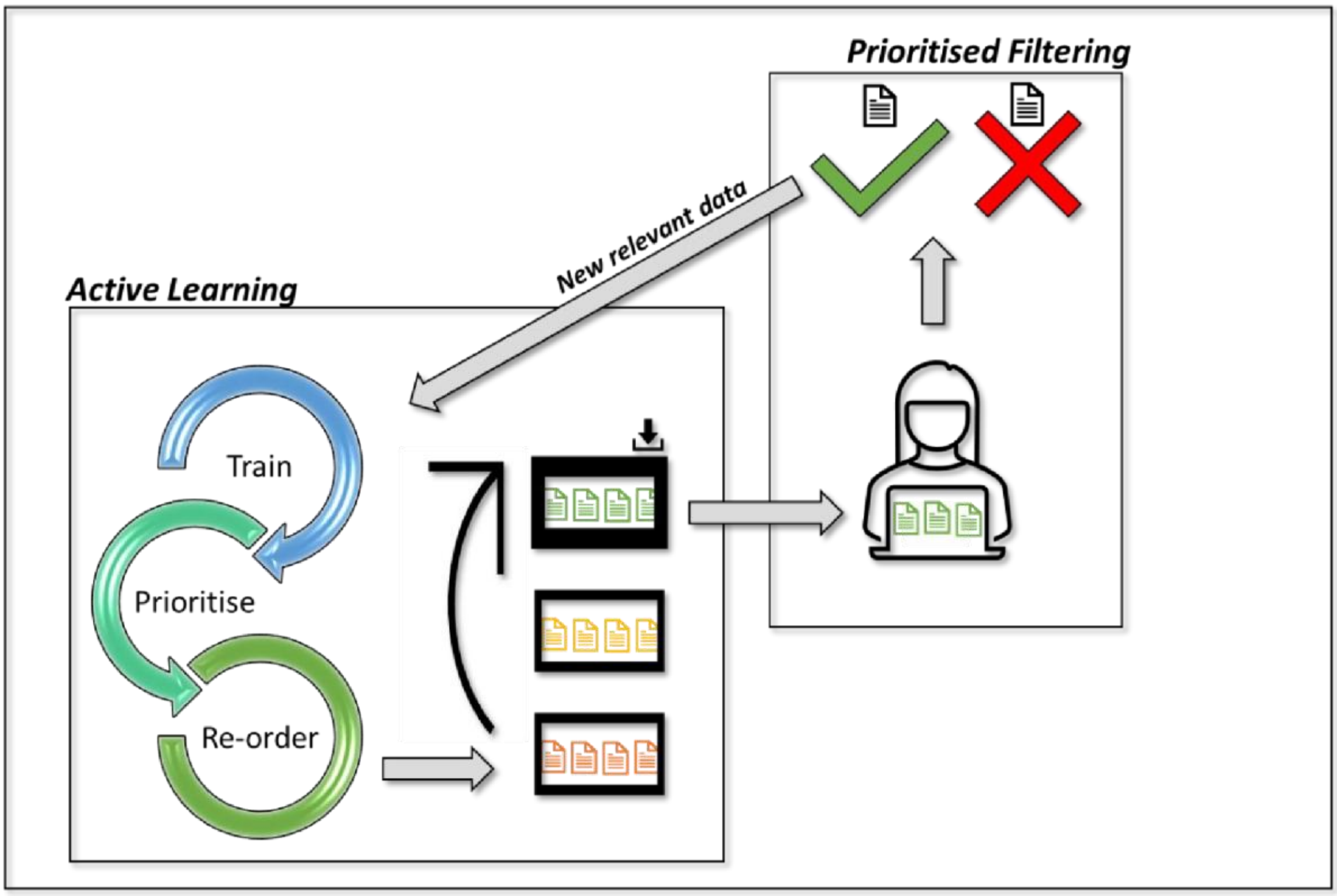

*Figure 3: Flow of information through the AIDOC tool Evaluation methodology*

For both tool outputs, we assessed the quality of reference re-ordering based on evaluation criteria established in the field of iterative learning for reference screening in systematic review automation. Since it is an iterative process, evaluation is commonly conducted retrospectively, by means of a simulation based on a fully manually screened gold-standard dataset with in- and exclusion decisions (Ferdinands et al., 2023; Singh et al., 2018). Recall, precision, and Work Saved over Sampling at 95% Recall are measures to quantify the theoretical potential of reduced screening effort. Datasets were labelled as part of real-world horizon scanning projects; manually by a single horizon scanning analyst. Whenever a horizon scan contained multiple data sources such as news articles, funding calls, trial registry entries, or peer-reviewed literature we conducted a separate analysis for each data type.

In the following, we explain the evaluation metrics used for evaluating reference ranking performance. The basic metrics are defined as follows:

- True Positives (TP) are references correctly classified as relevant
- True Negatives (TN) are references correctly classified as irrelevant
- False Positives (FP) are references that are irrelevant, but were classified as relevant.
- False Negatives (FN) are relevant references that were falsely classified as irrelevant (during iterative learning evaluation simulation, after each step of 'uncovering' prioritised labels, the unscreened relevant references are FNs).

**Recall**, also referred to as sensitivity, is the proportion of correctly-predicted relevant instances out of all relevant data points: TP/(TP+FN). It is commonly used to quantify the safety of AI methods where it is important that no relevant record is missed.

**Precision**, also referred to as positive predictive value, is the proportion of relevant records within all records predicted to be positive: TP/(TP+FP). It quantifies the amount of 'noise', or irrelevant data, that is contained in the model's positive predictions.

**WSS@r** (Work Saved over Sampling at target Recall) is a metric designed to quantify theoretical time-savings during screening at a desired recall threshold of x%, commonly 95% (Cohen et al., 2006). It is the percentage of references that would not require human review if screening were stopped at the target recall. WSS@r is corrected for the number of references that would have been expected to be found had the dataset been screened in a random order (Cohen et al., 2006). It is a useful retrospective evaluation tool but during a live (in-progress) project its results are not applicable because it is impossible to know the exact underlying recall in a partlyscreened dataset and thus difficult to justify a decision of stopping early. It was also shown to be affected by class-balance within a dataset, therefore Kusa et al. proposed a normalized version of this metric that is equivalent to the true-negative rate (Kusa, Lipani, et al., 2023).

**TNR@r** (True Negative Rate at target Recall) is the normalised equivalent of WSS@r as proposed by Kusa, Lipani, et al. (2023). It accounts for imbalanced distributions of positive ground truth labels and facilitates comparisons of classification performance across datasets.

We adapted open-source code[8] from Kusa for WSS@r and TNR@r calculation.

---

[8] https://github.com/WojciechKusa/normalised-precision-at-recall/blob/main/notebooks/evaluation.py (last accessed 19/03/2026)

We furthermore calculate the average precision and recalls at predefined cut-offs of 50 and 75% of the screening/filtration process and report the average percentage of records needing to be screened until the last included record is found in each dataset. We also calculated average precision and recall after screening 90 and 95% of all data but report these findings only in the appendix.

We simulated the iterative learning performance across 15 randomly shuffled runs. For each dataset and run we used 5 random included records at the start as positive seeds for the AI. For datasets with <30 total relevant labelled records, we used only 1 reference as positive seed. When conducting a comparison with the published literature, we furthermore tested using 15 and 1 positive seeds across benchmark datasets.

**Evaluation of LLM ensemble model:** We explored the added value of adding ensembled LLM predictions to the aforementioned iterative learning simulation-based environment. We directly feed LLM predictions created based on a structured template without further prompt-refinement to AIDOC as input variables (as described earlier) and calculate WSS@r, TNR@r, and all other metrics from the iterative learning simulation environment as before. In reality, tool users likely want adapt prompts and re-run LLM classification throughout a scan; or switch between LLM-supported prioritisation environment, the baseline AIDOC method, and classic machine-learning to increase reference yield. Current retrospective evaluation systems are not set up for this ‘adaptive’ and practice-oriented workflow. In the context of this paper, we therefore limit LLM-based classification to a single prompt applied at the beginning of the filtration process. The creation of this single prompt is guided by a template, and in the context of this paper, the prompt is created ‘blindly’ and not optimised or altered; to avoid the issue of dataset contamination.

## Stopping screening early

We tested the potential of the open-source Python library ‘Buscarpy’[9] implementing CMH (Callaghan & Müller-Hansen, 2020) to predict when analysts may stop sifting early. Using information about the order of in-and excludes from the iterative learning simulation, CMH predicts the confidence of reaching a user-specified recall threshold. This, in theory, enables users to stop reviewing data early in a prospective manner (i.e. in ongoing projects that haven’t been reviewed in full). There exist a range of algorithms and tools with similar functionality (Howard et al., 2020; Molinari & Esuli, 2024; Stevenson & Bin-Hezam, 2023), but CMH was chosen due to being a functioning open-source python package and it’s integration in a recent evaluation framework and validation study where it was shown to adhere to its recall target while producing conservative time-savings (Repke et al., 2026).

We retrospectively evaluate the algorithm’s safety in making an early-stopping decision at 95% estimated recall, noting instances where the true underlying recall was <1% lower than the predicted stopping point as minor errors, <5% lower as medium, and everything beyond that as major error. Across all 15 simulation runs for each dataset and each implementation of our ranking algorithm, we report the average percentage of records that wouldn't need to be screened anymore at the time of predicted stopping, together with the average true underlying recall at that point, as main outcomes. If, for example, in one simulation run the algorithm predicted 95% recall ($p<0.05$) after screening 75% of a dataset, a user who stops screening at this time wouldn't need to review 25% of their dataset. Had the user discovered 97 out of 100 relevant articles at that time, the true underlying recall would be 0.97 and CMH would have been safe to use in the specific scenario because it adhered to the pre-specified recall target of 95%. We furthermore experiment with three different (and previously not validated) bias parameter settings to explore the trade-off between work-savings and algorithm safety (Callaghan & Müller-Hansen, 2020). The bias parameters allow less conservative calculation of results, pushing the algorithm towards greater work savings while at the same time potentially increasing the number of errors (i.e. stopping early prematurely without reaching target recall). We report results across all simulation runs for bias =1 (reference implementation), and bias 1.5 and 2 (unvalidated). For the unvalidated settings we provide an overview of errors in cases where the algorithm missed its target for each parameter setting.

[9] https://buscarpy.readthedocs.io/en/latest/ (last accessed 19/03/2026)

# Results

## Dataset collection

Initial experiments were carried out between June – December 2024, with LLM-based predictions updated in 2026. We collected 12 datasets from previously completed horizon scans completed before and during this period. Their characteristics are displayed in Table 2, including dataset size, scope, text types and the specific time horizons for each scan. The Innovation Observatory defines three different time horizons for their scans:[10]

- Emerging horizon: Early, weak signals of innovative or disruptive ideas and emerging technologies. Evidence often comes from grey literature sources such as patents, press releases and news, funding calls, and early-phase trial registry entries.
- Transitional Horizon: Technologies transitioning from early clinical trial phases towards pre-regulatory status. Evidence can still include grey literature but also gradually shifts towards published and academic outputs, such as trial registry entries, data from websites, conference proceedings, and funding calls.
- Imminent Horizon: Technologies that are usually known to interest holders because they are approved or are undergoing technology appraisal and regulatory processes. Multiple layers of comparably high-quality evidence are available, including for example peer-reviewed and published literature and regulatory documents, as well as late-stage trial registry entries.

Inclusion decisions were made by single reviewers, as is customary for horizon scans. We additionally use the four OHAT/NIEHS (National Institute of Environmental Health Sciences) datasets first published by Howard et al. (2016). These reviews are systematic and mapping reviews with broad searches and inclusion criteria and multiple topics under review; similar to the heterogeneous nature of horizon scanning. By including these datasets, we enable a comparison of our iterative learning system to the published literature.

[10] Horizons are the Emerging Horizon (https://io.nihr.ac.uk/what-we-do/emerging-horizon/), Transitional Horizon (https://io.nihr.ac.uk/what-we-do/transitional-horizon/ ), and Imminent Horizon (https://io.nihr.ac.uk/what-we-do/imminent-horizon/ ). Last seen 19/03/2026

*Table 2: Datasets included in the analysis. For datasets denoted with * we truncated long text descriptions such as news article full text or funding call description to 2000 characters, which corresponds to 300-350 words as is commonly used for abstracts in scientific publications.*

| Name | Text type | Horizon | Scope | Fields | Total N | N included (%) |
|---|---|---|---|---|---|---|
| **AI Technologies** | Trial registry | Emerging | Broad | Titles | 1727 | 143 (8%) |
| **MRP** | Trial registry | Imminent | Focussed | Titles | 280 | 74 (26%) |
| **GenAI Technologies 1 Trials** | Trial registry * | Transitional | Broad | Titles, Abstracts | 1141 | 143 (12%) |
| **GenAI Technologies 2 UKRI** | Funding calls* | Transitional | Broad | Titles, Abstracts | 984 | 127 (13%) |
| **GenAI Technologies 3 NIHR** | Funding calls | Transitional | Broad | Titles, Abstracts | 116 | 37 (32%) |
| **Woman Health** | Funding calls* | Emerging, Transitional | Focussed | Titles, Abstracts | 677 | 5 (0.7%) |
| **MRC funding scan** | Funding calls* | All | Broad | Titles, Abstracts | 599 | 108 (18%) |
| **Rapid Biomed 1 Genetic Engineering** | Journal Articles | Transitional, Imminent | Focussed | Titles, Abstracts | 343 | 45 (13%) |
| **Rapid Biomed 2 TX delivery** | Journal Articles | Transitional, Imminent | Focussed | Titles, Abstracts | 352 | 53 (15%) |
| **Rapid Biomed 3 Tissues Devices** | Journal Articles | Transitional, Imminent | Focussed | Titles, Abstracts | 348 | 80 (23%) |
| **Quantum Technologies** | News* | All | Broad | Titles, truncated full text | 1262 | 254 (20%) (sensitivity analysis version: 168 includes) |
| **Health Screening** | News* | All | Broad | Titles, truncated full text | 261 | 57 (22%) |
| **OHAT SWIFT BPA** | Journal Articles | Not applicable | Focussed | Titles, Abstracts | 7700 | 111 (1.4%) |
| **OHAT SWIFT Fluoride** | Journal Articles | Not applicable | Focussed | Titles, Abstracts | 4479 | 51 (1.1%) |
| **OHAT SWIFT Transgenerational** | Journal Articles | Not applicable | Broad | Titles, Abstracts | 48632 | 765 (1.6%) |
| **OHAT SWIFT PFOA-PFOS** | Journal Articles | Not applicable | Focussed | Titles, Abstracts | 6328 | 95 (1.5%) |

## Quantitative evaluation

We calculated work-savings and recall at different thresholds across 15 randomly shuffled runs for each dataset and report average and standard deviation (SD) results across all runs. When possible, we used record titles and abstracts or abstract-length equivalent text for news, funding calls, and trial registry entries. The raw data for each run of each dataset, as well as the full set of calculated metrics for each run and plots with gain curves from each run and dataset are available in the appendix.

*Table 3: This table shows average and SD results of the AIDOC classifier on all datasets in our analysis, across 15 randomly shuffled runs using 5 random positive seed records where possible. For datasets with less than 30 included records we used only 1 positive seed. The average overall results are shown in the last row. *denotes datasets with insufficient information content, where humans are required to seek additional information during screening.*

| Name | WSS@95r | TNR@95r | Recall@50 % screened | Recall@75 % screened | Last included record found at |
|---|---|---|---|---|---|
| **AI Technologies** | 0.56 ± 0.03 | 0.66 ± 0.03 | 0.97 ± 0 | 1 ± 0 | 70% ± 5% |
| **MRP*** | 0.07 ± 0.02 | 0.15 ± 0.02 | 0.55 ± 0.02 | 0.78 ± 0.02 | 95 ± 2 |
| **GenAI Technologies 1 Trials** | 0.34 ± 0.04 | 0.44 ± 0.05 | 0.86 ± 0.05 | 0.99 ± 0.01 | 80 ± 4 |
| **GenAI Technologies 2 UKRI** | 0.22 ± 0.03 | 0.31 ± 0.03 | 0.83 ± 0.02 | 0.96 ± 0.01 | 97 ± 2 |
| **GenAI Technologies 3 NIHR** | 0.22 ± 0.04 | 0.39 ± 0.06 | 0.76 ± 0.06 | 0.96 ± 0.03 | 81 ± 5 |
| **Woman Health** | 0.56 ± 0.06 | 0.61 ± 0.06 | 1 ± 0 | 1 ± 0 | 39 ± 6 |
| **MRC funding scan** | 0.23 ± 0.03 | 0.33 ± 0.04 | 0.80 ± 0.02 | 96 ± 0.01 | 84 ± 3 |
| **Rapid Biomed 1 Genetic Engineering** | 0.50 ± 0.03 | 0.63 ± 0.03 | 0.97 ± 0.01 | 1 ± 0 | 53 ± 2 |
| **Rapid Biomed 2 TX delivery** | 0.35 ± 0.02 | 0.47 ± 0.02 | 0.82 ± 0.03 | 1 ± 0.0001 | 67 ± 4 |
| **Rapid Biomed 3 Tissues Devices** | 0.33 ± 0.01 | 0.48 ± 0.01 | 0.77 ± 0.02 | 1 ± 0 | 71 ± 3 |
| **Quantum Technologies** | 0.003± 0.01 | 0.05 ± 0.01 | 0.53 ± 0.05 | 0.76 ± 0.03 | 99 ± 0.2 |
| **Health Screening** | 0.35 ± 0.05 | 0.50 ± 0.07 | 0.88 ± 0.03 | 0.98 ± 0.01 | 91 ± 4 |
| **OHAT SWIFT BPA** | 0.73 ± 0.01 | 0.79 ± 0.01 | 1 ± 0 | 1 ± 0 | 37 ± 1 |
| **OHAT SWIFT Fluoride** | 0.89 ± 0.01 | 0.95 ± 0.01 | 1 ± 0 | 1 ± 0 | 24 ± 9 |
| **OHAT SWIFT Transgenerational** | 0.68 ± 0.01 | 0.74 ± 0.01 | 1 ± 0 | 1 ± 0 | 69 ± 3 |
| **OHAT SWIFT PFOA-PFOS** | 0.80 ± 0.01 | 0.86 ± 0.01 | 1 ± 0 | 1 ± 0 | 27± 5 |
| **Average ± SD (all)** | 0.42 ± 0.26 | 0.51 ± 0.25 | 0.85 ± 0.15 | 0.96 ± 0.08 | 67.60 ± 24.88 |

Table 3 shows results (average ± SD) for each dataset and aggregated results using the baseline AIDOC implementation that uses only cosine similarity of title+abstract embeddings of up to 10 randomly selected previously discovered included records in each simulation step.

*Table 4: Comparison of AIDOC, AIDOC + SGD, and AIDOC+SGD+LLM ensembles, and results reported by Howard et al. (2016). Here we report AIDOC results using either 1 or 5 random seed references across 15 runs.*

| | OHAT SWIFT BPA | OHAT SWIFT Fluoride | OHAT SWIFT Transgenerational | OHAT SWIFT PFOA-PFOS | Notes |
|---|---|---|---|---|---|
| **Howard et al., 2016** | **0.80 ± 0.01** | 0.86 ± 0.01 | **0.74 ± 0.003** | 0.83 ± 0.01 | Using 15 seed references |
| **AIDOC** | 0.73 ± 0.01 | 0.89 ± 0.01 | 0.58 ± 0.01 | 0.70 ± 0.01 | Using 5 seed references |
| **AIDOC + SGD** | 0.73 ± 0.02 | 0.85 ± 0.01 | 0.68 ± 0.01 | 0.80 ± 0.01 | Using 5 seed references |
| **AIDOC+SGD+LLM** | 0.77± 0.01 | **0.90 ± 0.01** | **0.74 ± 0.01** | **0.84 ± 0.01** | Using 1 seed reference |

Table 4 shows the performance of different versions of AIDOC, when used as ensemble with the classic machine-learning algorithm SGD as well as using LLM predictions to boost scores. It also shows how many initial seed references were used, demonstrating reduced needs for screening seeds across model versions. In comparison to the reference dataset, our WSS@95r for the basic form of AIDOC is lower in three datasets and  better in one, as seen in Table 4. With the largest dataset (Transgenerational), AIDOC's basic algorithm was outperformed by 13%. This may be because its underlying neural model only uses a random 10 positive labels each time when updating reference ranks; the dataset in question has 765 positive instances and classic machine-learning, as used originally (Howard et al., 2016), uses all available records plus negative training examples to make predictions. We therefore also tested the two ensemble models, the first of which is combining AIDOC with SGD machine-learning where every 5$^{th}$ re-ranking is carried out by the SGD model that considers all known positive labels at any step. As expected, this yielded an improved performance on the Transgenerational mapping review and PFOA datasets but was equal or inferior to the other two models for the more average-sized and more confined research-topic datasets.

We then tested the second ensemble approach, which includes LLM predictions from blindly created and unrefined prompts, added as values of either 0 or 1. This boosts priority ranking and led to an average of 0.812 WSS@95r across all datasets, compared with the previously published 0.807. While this improvement of results on average is marginal, it is important to point out the difference in requirements for positive starting seeds between the methods. For ‘fair’ comparison with Howard et al. (2016) we furthermore tested different numbers of random seeds, i.e. using 15 random included seeds as done with their model (as opposed the best-performing model’s default of 1). For the affected datasets, this meant that more than 10% of all included records are revealed to the algorithm ‘for free’. On the AIDOC ensemble, this led to WSS@95r improvements of 2% on BPA and PFOS-PFOA respectively, but as such a large unbiased sample of included records is unlikely available in reality we chose not to display these results in the table.

| Name | WSS@95r | TNR@95r | Recall@50 % screened | Recall@75 % screened | Last included record found at |
|---|---|---|---|---|---|
| **AI Technologies** | 0.76 ± 0.02 (+20%) | 0.88 ± 0.02 (+22%) | 0.99± 0.01 (+1%) | 1 ± 0.001 | 64% ± 7% (-6%) |
| **MRP*** | 0.03 ± 0.02 (-4%) | 0.1 ± 0.03 (-5%) | 0.53 ± 0.02 (+2%) | 0.76 ± 0.02 (+2%) | 97 % ± 1% (-2%) |
| **GenAI Technologies 1 Trials** | 42 ± 0.02 (+8%) | 0.55 ± 0.03 (+11%) | 0.94 ± 0.01 (+8%) | 0.99 ± 0.01 (+0%) | 84% ± 5% (+4%) |
| **GenAI Technologies 2 UKRI** | 0.29 ± 0.03 (+7%) | 0.38 ± 0.04 (+7%) | 0.84 ± 0.02 (+1%) | 0.97 ± 0.01 (+1%) | 90± 4% (-7%) |
| **GenAI Technologies 3 NIHR** | 0.26±0.05 (+4%) | 0.44 ± 0.08 (+5%) | 0.81 ± 0.02 (+5%) | 0.97 ± 0.01 (+1%) | 83 ± 5% (+2%) |
| **Woman Health** | 0.86 ± 0.01 (+30%) | 0.92±0.01 (+31%) | 1±0 (+0%) | 1±0 (+0%) | 8 ± 1% (-31%) |
| **MRC funding scan** | 0.28 ± 0.01 (+5%) | 0.39 ± 0.02 (+6%) | 0.81 ± 0.02 (+1%) | 0.98 ± 0.01 (+2%) | 87± 1% (+3%) |

| | | | | | | |
|---|---|---|---|---|---|---|
| **Rapid Biomed 1 Genetic Engineering** | 0.49 ± 0.03 (-1%) | 0.61 ± 0.03 (-2%) | | 0.97 ± 0.02 (+0%) | 0.98 ± 0.0 (-2%) | 83 ± 0.4% (+30%) |
| **Rapid Biomed 2 TX delivery** | 0. 22 ± 0.06 (-13%) | 0.31 ± 0.07 (-16%) | | 0.76 ± 0.05 (-6%) | 0.96 ± 0.01 (-4%) | 88 ± 6% (+13%) |
| **Rapid Biomed 3 Tissues Devices** | 0.3 ± 0.01 (-3%) | 0.44 ± 0.01 (-4%) | | 0.80 ± 0.02 (+3%) | 1 ± 0 | 70 ± 2% (-1%) |
| **Quantum Technologies** | 0.29 ± 0.02 (+28%) | 0.42 ± 0.02 (+41%) | | 0.88 ± 0.01 (+ 35%) | 0.97 ± 0.01 (+21%) | 95 ± 1.5 (-4%) |
| **Health Screening** | 0.49 ± 0.02 (+14%) | 0.68 ± 0.03 (+18%) | | 0.96 ± 0.01 (+8%) | 0.99 ± 0.01 (+1%) | 66 ± 9 (-25%) |
| **OHAT SWIFT BPA** | 0.78 ± 0.01 (+5%) | 0.84 ± 0.01 | | 1±0 (+0%) | 1±0 (+0%) | 29± 2 |
| **OHAT SWIFT Fluoride** | 90 ± 0.01 (+1%) | 0.96 ± 0.01 | | 1±0 (+0%) | 1±0 (+0%) | 15 ± 1 |

| | | | | | | | |
|---|---|---|---|---|---|---|---|
| **OHAT SWIFT Transgenerational** | 0.74± 0.01 (+6%) | 0.8±0.01 (+6%) | | | 1±0 (+0%) | 1±0 (+0%) | 63 ± 4 |
| **OHAT SWIFT PFOA-PFOS** | 0.84 ± 0.01 (+4%) | 0.91 ± 0.01 | | | 1±0 (+0%) | 1±0 (+0%) | 14 ± 1 |
| **Average ± SD (all)** | 0.50 ± 0.27 (+4%) | 0.60 ± 0.25 (+5%) | | | 0.89 ± 0.12 (+2%) | 0.97 ± 0.05 (+1%) | 64 ± 0.29 (-3%) |

We also conducted a ‘stress-test’ on the remaining models, namely of providing only one random starting seed but observed that AIDOC’s results stayed consistent throughout, showing <1% variation of averaged results and no increases in SD revealing less included records at the start. We therefore decided to report the results with 5 or 1 seed respectively, as these are more representative of real use-case scenarios.

*Table 5: Simulation results using the AIDOC ensemble classification (SGD+LLM). Numbers are displayed as averages ± SD across 15 runs, using 1 random included record as seed at the start. The number displayed in parentheses is the performance difference to the baseline AIDOC classifier displayed in results table 3.*
**denotes datasets with insufficient information content, where humans are required to seek additional information during screening*

Finally, we applied the AIDOC SGD+LLM ensemble model to all datasets, and computed score differences between the approaches. On average, using the blindly created prompts guided by the prompt template, a 5% increase in TNR@95r was observed. Performance gains were mostly observed during the initial stages of screening, with underlying recall at 50% and 75% of screening only improving by 2% and 1% respectively.

The evaluation on theoretical, normalised work savings across datasets shows that on average, 60% filtration effort could be saved at the underlying 95% recall level when using the true-negative rate, which is also called normalised WSS (50% using older WSS@95 metric). It should be noted that these results include one dataset where the textual information did not include sufficient information for decision making, and that excluding this dataset would have led to increased work-savings of 5% in both metrics.

When evaluating recall at a pre-specified cutoff of manually assessing 50 and 75% of each relevant dataset, recalls reached 0.92 (± 0.09) and 0.99 (± 0.01) on average. This means that when using AIDOC to prioritise references during filtration on datasets with characteristics comparable to ours, horizon scanning analysts may find on average 99% (± 1%) of all relevant records after reading the first 75%. For high priority rapid scans, they may find on average 92% of all includes after assessing half of the dataset, although higher SD of 9% means that the results at this threshold are less stable.

One dataset (MRP trial registry entries) is included data that were not adequate to inform the AIDOC classifiers and thus results were equal to presenting references in random order. Another dataset (Quantum Technologies) is used in a sensitivity analysis exploring impact of in-or excluding data of different time horizons.

For MRP, analysts needed to consult sources on rare diseases beyond the dataset to reach inclusion decisions, which explains the poor performance of AIDOC on the available data only This negative finding raises important practical considerations for the usage of AI in 'real-world' horizon scanning and is further discussed in the following section.

For Quantum Technologies, the dataset included long news article texts with descriptions of a very distant and unspecific horizon; and technology that may be promising but, in many ways, has not been applied in healthcare yet (i.e. hypothetical or speculative information on likely developments that may only be beneficial in health contexts in the future). We did, therefore, decide to experiment further with the quantum dataset to a) explore impact of including LLM inclusion predictions in the process, and b) re-label its included records with stricter adherence to inclusion criteria of the application of quantum technology in healthcare (i.e. removing the emerging horizon and indirectly relevant articles from the included records).

**Error! Reference source not found.** shows results of this on our SCANAR-derived news datasets for health screening and quantum technology (both original and relabelled versions for quantum data). The addition of LLM alone led to a performance increase of 41% TNR@95% recall and the ability to find 97% of all included articles after screening 75% (see **Error! Reference source not found.** and Figure 4 a+b). On the re-labelled dataset, we explored the impact of excluding the vague and distant horizon includes (sensitivity analysis) which showed that these records indeed contributed the most errors within the simulation study (see **Error! Reference source not found.** and Figure 4 c).

*Table 6: LLM + ML+ AIDOC ensemble approach using template-guided prompts and full texts of news articles retrieved by SCANAR. The numbers in parentheses indicate performance gains over the AIDOC baseline approach. On the original labels for both datasets, the LLM approach showed clear work-savings. On the edited version of the Quantum Technology scan where we excluded articles of the Emerging Horizon, work-savings increased further.*

| Name | WSS@95r | TNR@95r | Recall@50% screened | Recall@75% screened | Last included record found at | Note |
|---|---|---|---|---|---|---|
| **Health Screening LLM + ML + +AIDOC** | 0.49 ± 0.02 (+14%) | 0.68 ± 0.03 (+18%) | 0.96 ± 0.01 (+8%) | 0.99 ± 0.01 (+1%) | 66 ± 9 (-25%) | Using full text |
| **Quantum Technology LLM+ ML + AIDOC** | 0.29 ± 0.02 (+28%) | 0.42 ± 0.02 (+41%) | 0.88 ± 0.01 (+35%) | 0.97 ± 0.01 (+21%) | 95 ± 1.5 (-4%) | Using full text, full original dataset |
| **Quantum Technology (imminent/transient) LLM + AIDOC** | 0.50 (± 0.04) (+49%) | 0.63 (± 0.05) (+62%) | 0.96 ± 0.01 (+43%) | 0.99 ± <0.01 (+24%) | 70 ± 3 (-29%) | Sensitivity/ Error analysis |

## Simulating early stopping

Figure 5 shows results of applying the CMH early stopping algorithm on the results from our iterative learning simulation study on news articles. In Figure 5a) we tested three bias parameters and their impact on work-savings for the re-labelled Quantum dataset and found CMH to be 100% safe using default parameters. Increasing bias to

1.5 led to 10% additional work-savings, but also one single simulation run (out of 15) where the algorithm predicted 95% recall while the true underlying recall was marginally lower, at 0.946. Using bias parameter 2, work-savings further increased, and the marginally lower-than-estimated recall runs increased to two, which may be acceptable in a horizon-scanning scenario, depending on the context of the scan.

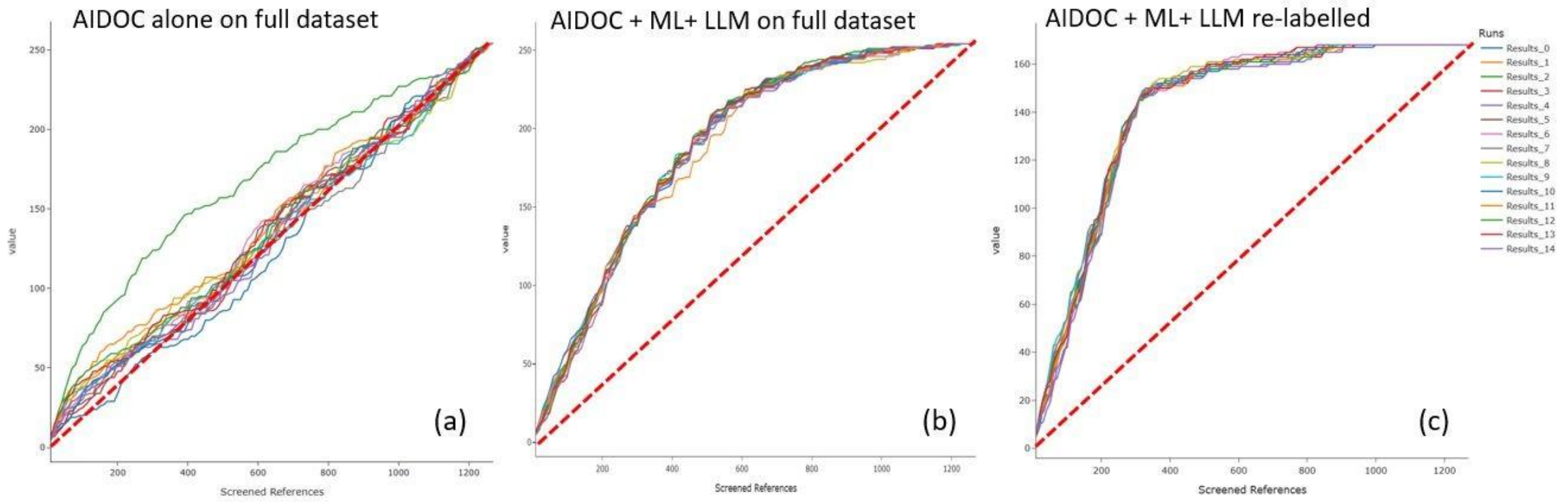

*Figure 4: Performance of reference priorisation on the challenging SCANAR-derived Quantum Technology dataset, containing full text news articles across multiple horizons. The red dashed line indicates how well random sampling of articles would have performed in comparison to the AI-supported workflow. a) Shows poor performance of AIDOC alone. b) Uses LLM labels within the ensemble model to prioritise LLM-included articles. c) Re-labelled dataset, excluding vague and distant articles to visualise that they contributed most errors.*

Figure 5b) shows that CMH found safe stopping thresholds with all tested parameters on the Health Screening dataset, with work-savings up to 27%.

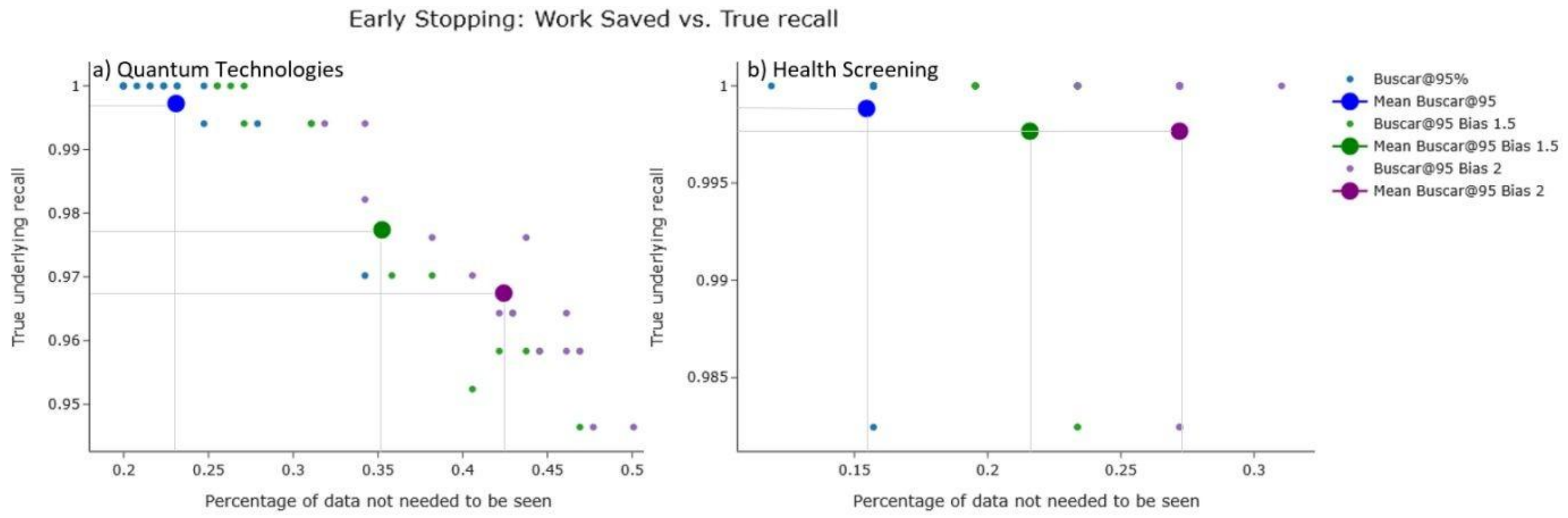

*Figure 5: Results of applying CMH on the results from the iterative learning simulation study on news articles. We set it up to predict when we may safely stop screening early at 95% estimated recall, and show work savings in terms of the percentage of articles that wouldn't need to be seen on the X axis for each simulation run. They Y axis shows true underlying recall for each predicted stop of screening. Plot a) We tested three bias parameters and their impact on work-savings for the re-labelled Quantum dataset. Plot b) CMH found safe stopping thresholds with all tested parameters on the Health Screening dataset, with work-savings up to 27%.*

We then applied CMH to the simulation results of all 16 datasets, to determine its ability to provide a reliable prediction of reaching a target recall of 95% with $p < 0.05$. The algorithm was applied with it's more conservative and validated bias parameter of 1, in addition to experimental usage of using 1.5 and 2.

Full results per dataset, on each randomly shuffled run, per bias parameter are shown in the appendix, for the baseline classifier and for the LLM ensemble method.

When applied to simulation results from the AIDOC baseline classifier, it would have, in theory, enabled reviewers to avoid assessing 17% of their data (bias 1). When increasing the bias parameters, the percentage of data not needing to be assessed rose to 23 and 27% respectively. However, this also led to an increase in errors that centered around the Quantum Technologies datasets (see Appendix for full data).

Figure 6 shows performance on the results of different CMH configurations applied to outputs of the AIDOC LLM ensemble classifier. As before, CMH at bias 1 met its target recall across datasets and would have saved users from screening on average 20% of their data (+3% increase). Usage of the higher unvalidated bias parameters consistently led to higher work-savings, with 27% and 32% of datasets not needing to be reviewed manually when stopping early respectively. This did, however, produce a small amount of results where target recall was not met in single runs. On average, target recall was met for each bias parameter on each dataset (see full data in Appendix).

The more difficult MRP and Quantum datasets each produced one run with 0.94 true underlying recall in the bias 1.5 scenario. On bias 2.0, MRP produced one run with true underlying recall of 0.93 and five with 0.94. Two other datasets (GenAI Technologies 2, Rapid Biomed 2) also produced minor errors with the high bias parameter) but on average, CMH met its expected recall.

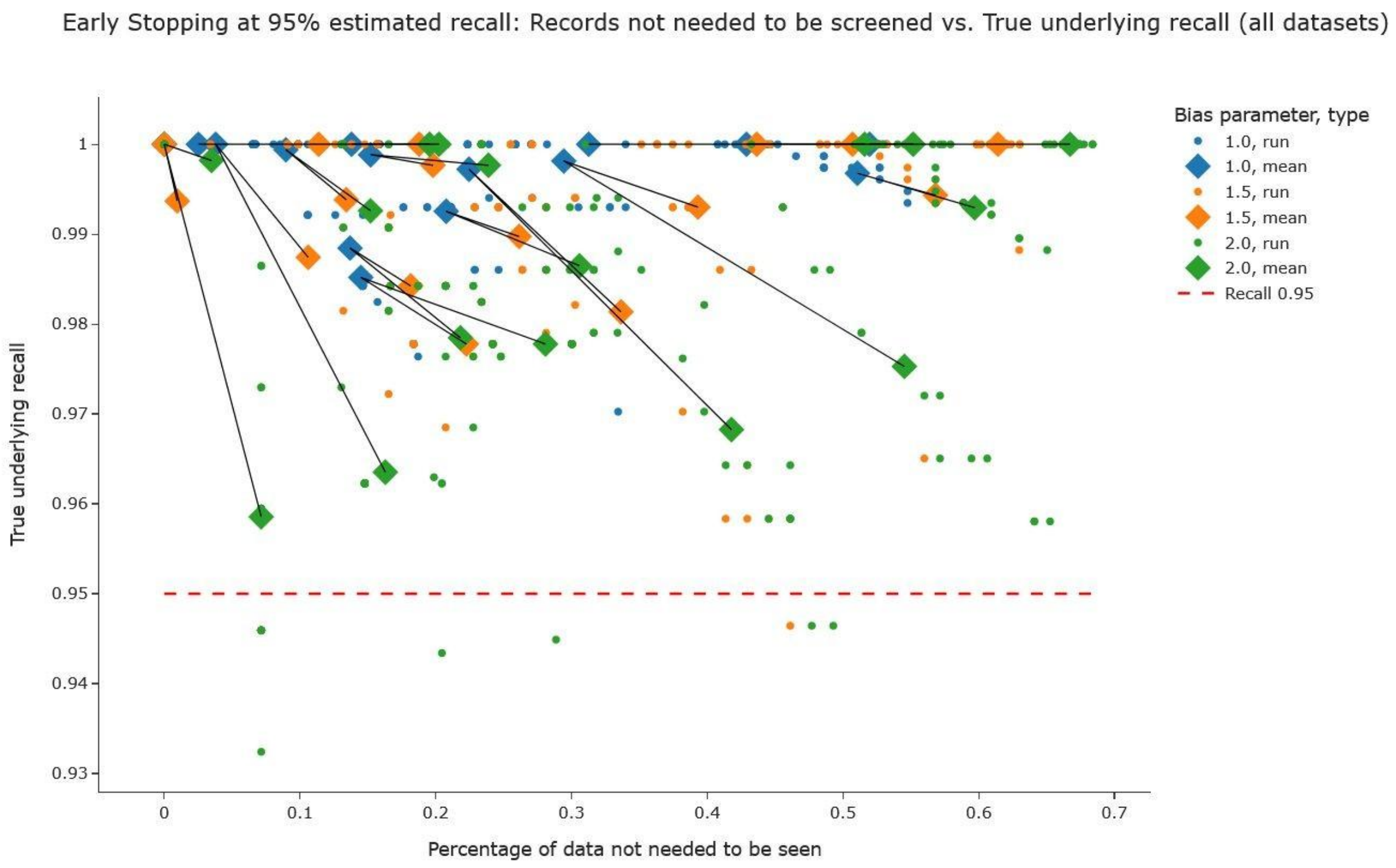

*Figure 6 Results of applying CMH on the LLM ensemble classifier results. The figure displays results from all datasets across three bias parameters. Diamonds indicate the average result for a dataset, while small circles indicate exact results from runs. Average results for each dataset are connected by black lines, to show impact of changing bias parameters. The red line indicates the target recall threshold, and circles underneath this line represent instances where the algorithm predicted the stopping point too early.*

## Discussion

Research into machine learning and automation tools aimed at enhancing efficiencies in evidence synthesis, specifically in systematic reviews, has become a wellestablished field (Johnson et al., 2022). However, this progress has not yet extended to horizon-scanning or other methodologies that necessitate the use of soft intelligence such as news and other grey literature. Specifically, the identification and retrieval of relevant results and the extraction of data are posing challenges in medicines and
MedTech horizon-scanning, given the rapidly evolving landscape and vast amounts of data from online sources. In the following sections we discuss implications in practice of using the two tools described in this manuscript.

## Implications in practice

### *News retrieval using SCANAR*

Over the course of 1 year we applied SCANAR to retrieve news articles for a variety of scans in the NIHR Innovation Observatory, including topics on Alzheimer's disease and diagnostics, traumatic brain injury (Oparah et al., 2023), quantum technology (Fairbairn et al., 2025; Sharma et al., 2025), metabolic hormones, AI in healthcare (Lanyi et al., 2025), and health screening. Before SCANAR, news scanning workflows involved triaging and manual export/deduplication of results in a spreadsheet; a process that was repeated for each query. This approach was time-consuming and limited the scope, transparency, and reproducibility of news scans.

SCANAR increases the scope of news scans via automated bulk-retrieval of 100 news items per search query, allowing as many queries as needed to systematically retrieve information and store it in a standardised spreadsheet. The tool increases reproducibility by retrieving all articles and their full texts; as opposed to the manual workflow that only catalogued articles deemed relevant by a single analyst. This allows users for example to revisit results of a previous scan if a technology was missed and to track which query produced the most and the least relevant articles. Another factor that increases reproducibility is that the tool can be deployed on a server in the that does not retrieve results tailored to personal search histories and browser cookies, and it enables automation of 'living' news scans, such as one weekly updated scan about health screening that entered it's second 'living' year in 2026. For transparency, SCANAR records the number of records retrieved for each query, the number of duplicate articles between the queries, as well as the publication date and news outlet for each article. Despite its straightforward core functionality of querying the Google News RSS API, the batch-search, deduplication, ranking, export functions, and userinterface add value to horizon scanning in practice. This value is generated mostly by increasing the reproducibility, data collection ability, and traceability of results. Example SCANAR output for the Quantum Technology scan is attached as appendix.

### *Semi-automation of screening using AIDOC and LLMs*

For screening automation, we used datasets of previously completed scans, as well as news datasets derived from the SCANAR tool to explore further time-saving potential for news scans. We also compared the results of our classifiers with those reported by Howard et al. (2016) in their publication that averages the best performance across the commonly used Cohen systematic reviews dataset (Cohen et al., 2006; Howard et al., 2016) when normalised and independently compared with five other classifiers (Kusa, Lipani, et al., 2023). We computed comparative scores between the best-performing model and ours for a smaller, more heterogeneous dataset that includes mapping and scoping reviews in the domain of environmental toxicology. Our results were comparable to this state-of-the-art published classifier, outperforming it on one dataset using AIDOC, and attaining superior performance when integrating LLMs into the retrospective simulation study. Most notably, our main reported scores were generated using only 1 or 5 random included references revealed as positive screening seeds at the beginning of the simulation process across the comparative datasets, and reproduced with a <1% variance of results when using only 1 seed to start. Howard et al. (2016) reported using 15 labelled relevant records at the start of their simulation – thus revealing a much larger proportion of the ground truth to the model 'for free'.

In practice, the main drawback of using traditional ML approaches is this requirement of a higher number of included records to gain efficiencies in the early screening process. When applying iterative learning in practice in horizon scanning, it would be challenging to provide a balanced and at the same time unbiased set of 15 (or even 10) seed records at the start. The same holds true for systematic reviews; for many of which 15 relevant records may make up a significant amount of all available evidence. Most relevant evidence is unknown at the start of the screening process and thus results from the published automation literature that use algorithms with 15 known starting references may overestimate the downstream 'real-life' work savings of any AI algorithm. This holds true especially if reviewers are forced to do an initial spell of random reference sampling and screening until enough training records are uncovered.

By starting with 1 or 5 known records without impacting overall results (and only ever utilising a maximum of 10 randomly sampled relevant records at each iterative reference re-ranking), the approach presented in this manuscript reduces the reliance on prior knowledge of relevant evidence at the start of reference filtration and our additional experiments integrating LLMs show that the previously limiting requirement for prior knowledge may be reduced even further.

In our evaluation, efficiency of AIDOC was shown especially for datasets with a low number of includes and tighter topic focus. For broad mapping reviews with a high ratio of includes we introduced an ensemble algorithm combining AIDOC with SGD to utilize the strengths of both approaches.

## Prospective usage of automation in screening

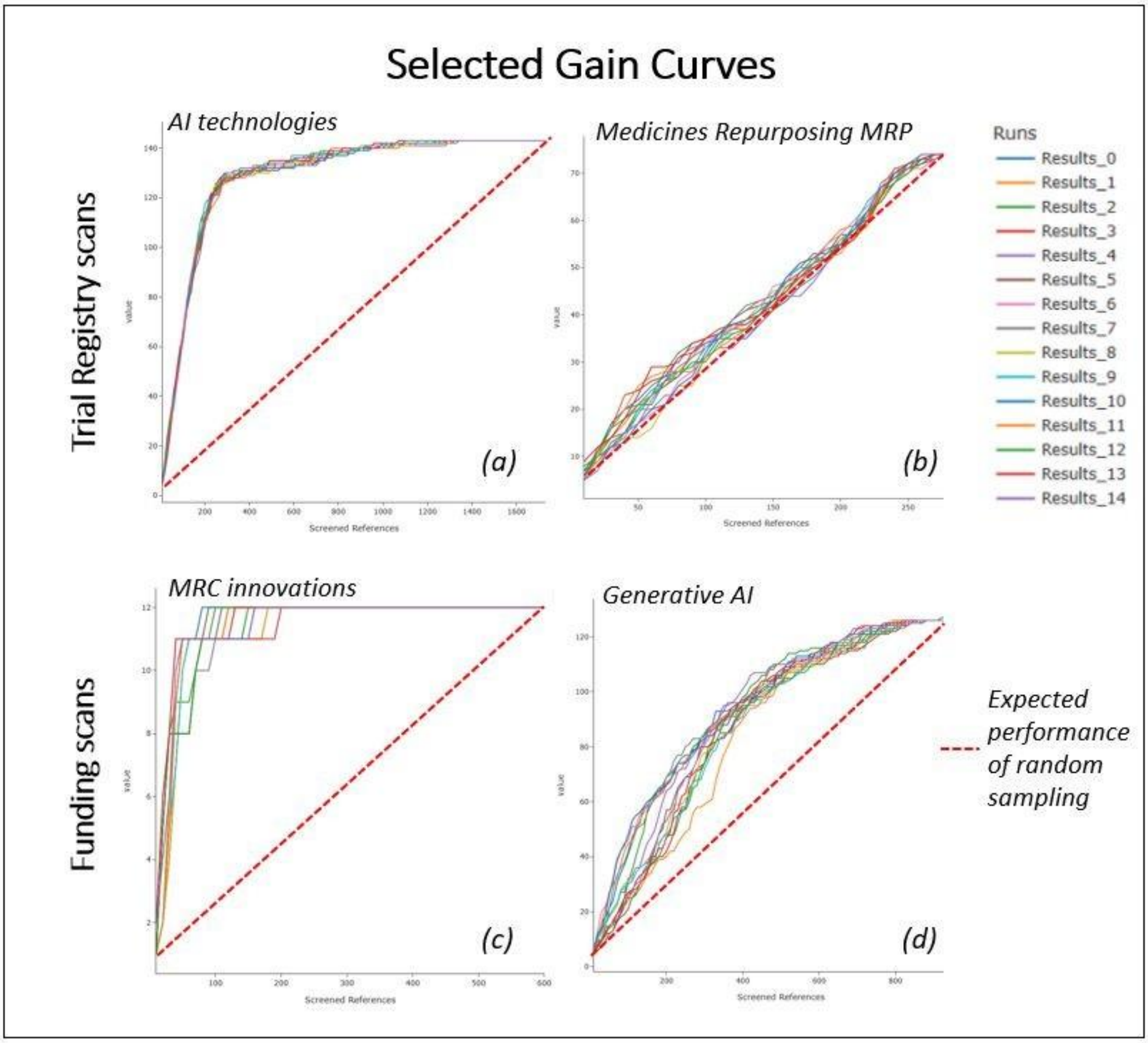

*Figure 7: Gain curves, showing screening simulations across 15 random runs for 5 exemplary datasets, with comparison to the screening progress expected for random reference sampling. (a) Clinical trial registry entries about AI innovations on a dataset with good performance; (b) Clinical trial registry entries about rare and repurposed medicines showing no better than random sampling performance due to insufficient information within dataset; (c) Funding scan with MRC records showing good performance; (d) Funding scan on generative AI innovations showing below average but better than random performance.*

Choosing to reveal only 1 random include and simulating the screening process, the results shown in the previous section demonstrate that 65% theoretic work savings were attainable using the AIDOC ensemble classifier that includes LLM, or 62% worksavings without using the additional model. Using the external data-driven stopping algorithm CMH with its validated bias parameter 1 (Callaghan & Müller-Hansen, 2020), we demonstrated that more safe thresholds leading to reduction of 20% workload were possible.

We calculated these numbers using heterogeneous datasets, including two that included inadequate text information for the classifiers to work with. For these scans, analysts were for example required to visit additional websites and to collect external data not reflected in the search results or texts were too long. Thus the tool was not able to provide work savings above and beyond random sampling of references. It is important to keep in mind the heterogeneity of horizon scanning work and to remind prospective users of automation tools to be aware of the data (quality) fed to an AI model and to recognise situations where the AI clearly provides no added value. Here, the responsibility lies with tool developers to include visualisations of gain curves and an in-progress evaluation based on the user's labelled data to alert them when reference priorisation is failing.

Examples to illustrate how gain curves can be used as visual cues are shown in Figure 7. When updating a gain curve of the actual 'live' screening or filtration process in conjunction with the diagonal line representing random sampling, users will be able to quickly identify if AI priorisation is providing extra value beyond random sampling.

If the gain curve crosses the diagonal line, as seen in Figure 7b, then the users should be required to review all records in the dataset manually. If the gain curve does not show the typical 'Knee' shape (as seen in 7a and c) but rather runs in proximity to the diagonal line, users should exercise caution and carefully weigh the risk of missing records against the resources they are able to spend on their horizon scan. In any case, we suggest that scans using reference priorisation and potential early stopping of filtration should provide a mini-report containing their gain curve and WSS@95, nWSS@95, recall at 50% and at 75% based on the data they have screened in any project (even if they did not assess the full dataset). AIDOC provides a functionality to the users where they may treat their viewed records as a 'complete' dataset in order to create the output. Naturally there is uncertainty in results computed based on incompletely labelled datasets. However, the purpose of reporting these results is not to quantify AI performance with the aim to compare its performance externally, but rather to provide transparency of the process to the commissioners and interest holders of a horizon scan. It serves as a sanity check that the AI can learn from data of sufficient quality to add value to the filtration process.

In horizon scanning, currently there are no methodological guidelines that recommend thresholds for the sensitivity and specificity of searches. Given the lack of reporting methods in general in this field, it is not known what sort of trade-offs between rapidity and capacity are influencing the conclusions of scans that do not use automation at present (Garcia Gonzalez-Moral et al., 2023). Our results have shown that SCANAR can alleviate challenges in information retrieval by searching and aggregating news articles automatically, eliminating previous human workload of conducting searches and manually exporting and deduplicating results. AIDOC has a related potential to alleviate the impact of methodological shortcuts in information retrieval for urgent scans – allowing more sensitive searches and enabling analysts to identify on average 92 and 99% of all relevant records after filtering through 50% and 75% of the data.

## Future research

Research into a reliable threshold for terminating filtration early requires further investigation into existing data-driven (statistical) stopping algorithms and their applicability to heterogeneous horizon scanning datasets beyond the one tested method in this article, for example in the scope of a more comprehensive analysis as performed by Repke et al. (2026). Typically, the evaluation of stopping criteria is based on dual-screened datasets from systematic literature reviews and peer-reviewed data, which may limit the generalizability of published results to our research topic.

Secondly, we aim to further explore the value of different methods to integrate LLMs to automatically assign in- and exclusion labels. This paper includes datasets with challenging data for existing neural and machine-learning methods. We demonstrated that between 18-41% additional improvements in work-savings may be attained on the most challenging datasets, but more in-depth research is clearly warranted. The larger context windows of LLMs as well as their capacity of applying general knowledge are the likely cause for the improved results. However, there may be several methods of developing and integrating LLM prompts and one challenge for further work is to explore workflows that make LLMs safe for usage in prospective scenarios (rather than retrospectively quantifying their performance based on a labelled dataset). The quality of LLM output is influenced by the quality of a user's prompts, and might be limited by other factors such as context length, parameter settings, task complexity, knowledge cutoff, model provider's system prompts and safety guardrails, and model version. This adds practical challenges that are not addressed in current retrospective validation workflows.

Additionally, more use-case evaluations for horizon scan automation are required. Although our tools have the potential to accelerate the horizon-scanning process, thereby enabling more efficient resource management and delivering more timely results to interest holders and the public, further empirical assessments are essential to validate their effectiveness in less-controlled scenarios.

## Limitations of the tools

Upon examining the results and the selected gain curves, it becomes evident that the effectiveness of the tools is contingent upon the specific characteristics of the horizon scan to which they are applied. For example, AIDOC alone demonstrates reduced efficacy in the MRP and Quantum scans. This is likely attributable to the fact that the titles of news sources do not provide sufficient information to be directly included, and the full texts are extensive, including substantial amount of information within niche and nuanced topic areas. The MRP scan looked at repurposed medicines for specific indications, requiring an extra workflow step of identifying each repurposed medicine against a secondary database. Consequently, it went beyond the scope of the automation tool  to identify scientifically relevant data from these results as it did not have the necessary secondary information. Additionally, scans that have a higher number of overall includes (such as the Quantum scan), require more manual review by topic expert researchers.

Another limitation to include is the span of the time horizons. The Quantum scan was exploratory, searched more distant and emerging time horizons This resulted in data that are more difficult to comb through even for humans to find relevant information, and analysts being over-inclusive. Scans looking at the imminent or transitional horizon have more regulatory and clinical information, providing more consistent terms and data that the tool can prioritise with more accuracy. In future work we aim to assess the performance of different classification systems that may be able to address these limitations, including for example agentic workflows that gather additional context information about a technology or medicine that appears in a record.

Regarding AIDOC, one limitation of the current system is that it does not offer a full user-interface for screening, but rather allows users to upload spreadsheets with data and to download a re-ordered version.

Regarding the implementation of LLM-based screening, a limitation of the current work is that we formulated only one prompt for each dataset, guided by the structured template shown in Table 1. We drew information from the inclusion criteria of the underlying project of each dataset, from a protocol if possible. We could have split the gold-standard datasets into prompt development and test datasets, which is generally considered best practice for binary classification scenarios, and reported results for the development set separately. However, for this proof-of-concept we were not interested in optimising a prompt based on labelled data that realistically will not be available at the start of a fresh horizon scanning project anyway. We were more interested in observing if the LLM can add value to an iterative learning process. Therefore, we adopted a ‘blind’ one-shot only approach guided by the prompt template, feeding a one-off set of LLM predictions into the iterative learning simulation. By developing and applying the one-off prompt blindly we avoid the negative impacts of overfitting, and at the same time realistically simulate how adding LLMs to the workflow using template guidance may influence results from the start of a horizon scan. As mentioned before, more in-depth research of how to integrate LLM filtration for horizon scans and reference screening in systematic reviews in practice is warranted - but not scope of this current paper.

## Conclusion

The challenge of retrieving and assessing an overwhelming amount of data is a commonly seen across horizon scans. Horizon scans typically have more rapid turnaround times compared to traditional systematic reviews. However, the increasing volume of information online has led interest holders to demand even more rapid, broader, and frequent scans. This requires the development and implementation of bespoke tools for horizon scanning and evidence synthesis to manage the growing data efficiently and meet the evolving needs of interest holders, particularly within the healthcare sector. In this paper we present, validate and discuss the first two modular tools that will eventually form the Innovation Observatory's Horizon scanning toolkit. This toolkit will aim to encourage standardised workflows for emerging, transitional, and imminent horizons. It will include a mixture of AI-free tools that automate processes, such as SCANAR, and human-in-the-loop AI-support tools, such as AIDOC, that integrate machine-learning as well as generative and non-generative large language models.

## Acknowledgements

We would like to thank our colleagues from the NIHR Innovation Observatory and Newcastle University for their contributions to this work. Saleh Mohamed for help with AWS deployment; James Woltman, Claire Eastaugh, Dawn Craig, Jaume Bacardit, Nick Meader for input to the design phase of tools; Kate Lanyi, Anjum Jahan, Pauline Addis, Megan Fairweather, Andrew Mkwashi, and Sola Akinbolade who supplied datasets as well as all analysts who carried out the primary filtration work on which we based our retrospective evaluation.

Furthermore, we would like to thank Katy Town, Omaer Syed and Rebecca Dliwayo from the UK National Screening Committee's Horizon Scanning team for their feedback and for sharing dataset labels with us.

## Funding

This project is funded by the National Institute for Health and Care Research (NIHR) [HSRIC-2016-10009/Innovation Observatory]. The views expressed are those of the author(s) and not necessarily those of the NIHR or the Department of Health and Social Care.

## Data Availability

All code for the tools described in this article can be found on GitHub. The tools themselves as Streamlit app code are here (https://github.com/L-ENA/SiftingApps/tree/main ). More detailed code for the iterative learning , LLM, and early stopping experiments including their evaluation is here: (https://github.com/L-

ENA/SiftingSimulations). Datasets, prompts, analysis data, including the predicted order of references for each dataset on each run, full results for each run, gain curves for each simulation, and plots showing CMH results on 3 bias parameter settings are in our OSF repository: https://osf.io/qapwe/

Here is a description of appendix files and their content:

- AIDOC_results_per_run.zip: calculated results for each dataset
- AIDOC_predictions_per_run.zip: results for random runs for each dataset, also called qrels
- AIDOC_gaincurves.zip: gaincurve visualisations for each dataset
- AIDOC_buscar.zip: CMH stopping visualisations for each dataset
- CMH_results_summary_baseline.xlsx: CMH per run results on each dataset and each random run
- CMH_results_summary_ensemble.xlsx: CMH per run results on each dataset and each random run
- datasets.zip: files containing the datasets and labels
- ensemble_buscar.zip: CMH stopping visualisations for each dataset
- ensemble_gaincurves.zip: gaincurve visualisations for each dataset
- ensemble_global.zip: calculated results for each dataset
- ensemble_runs.zip: results for random runs for each dataset, also called qrels

- managed_runs.json: LLM prompts and column mappings for each dataset

## Conflict of Interest

All authors: None declared.

## Credit statement

**Conceptualization**: LS, OS, CM, SGGM

**Methodology, Software, Validation, Formal Analysis**: LS

**Writing - Original Draft**: LS, OS, SGGM

**Writing - Review & Editing**: LS, OS, CM, SGGM

**Visualization, Supervision**: CM, SGGM

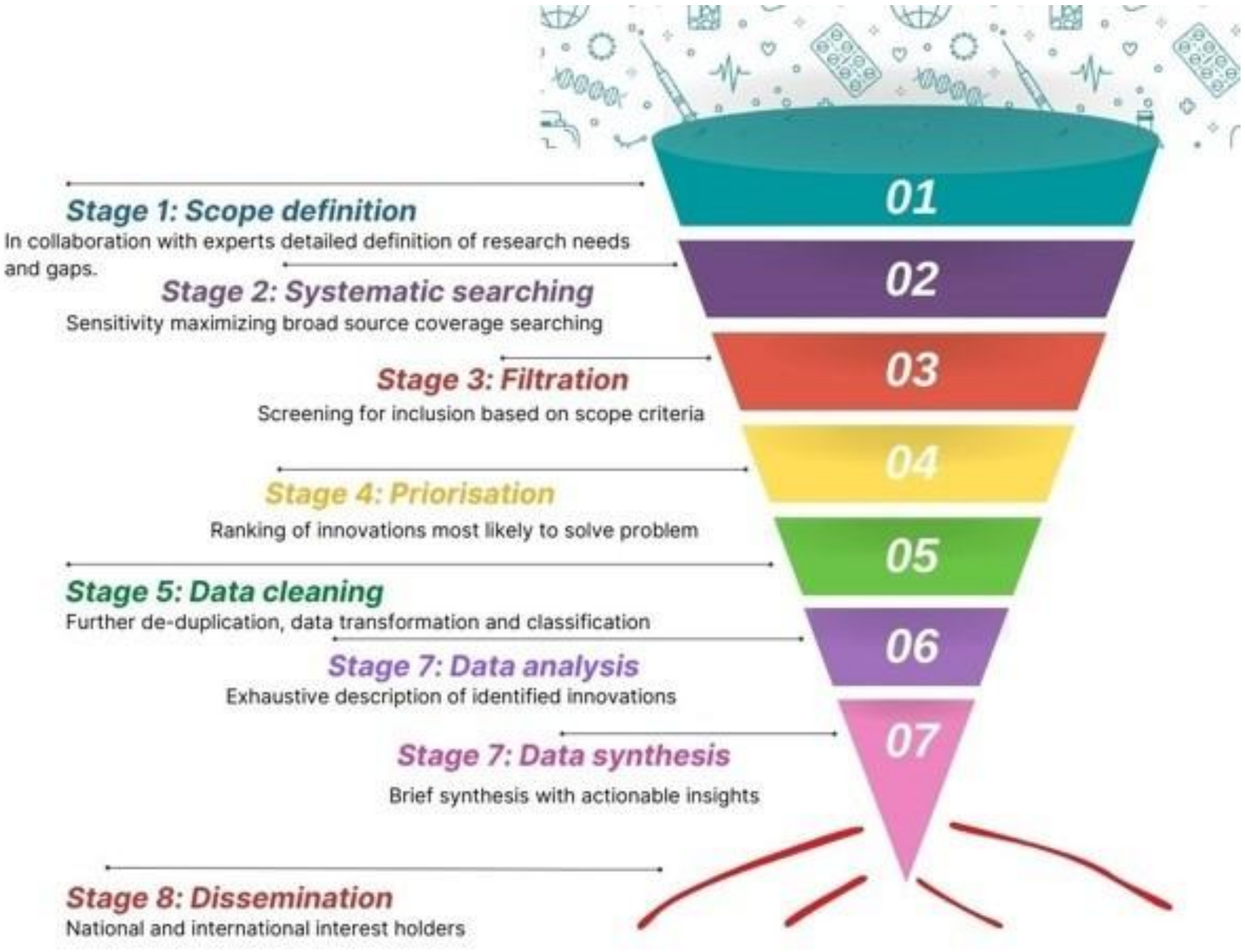

Figure 1: Horizon scanning stages

159x119mm (96 x 96 DPI)

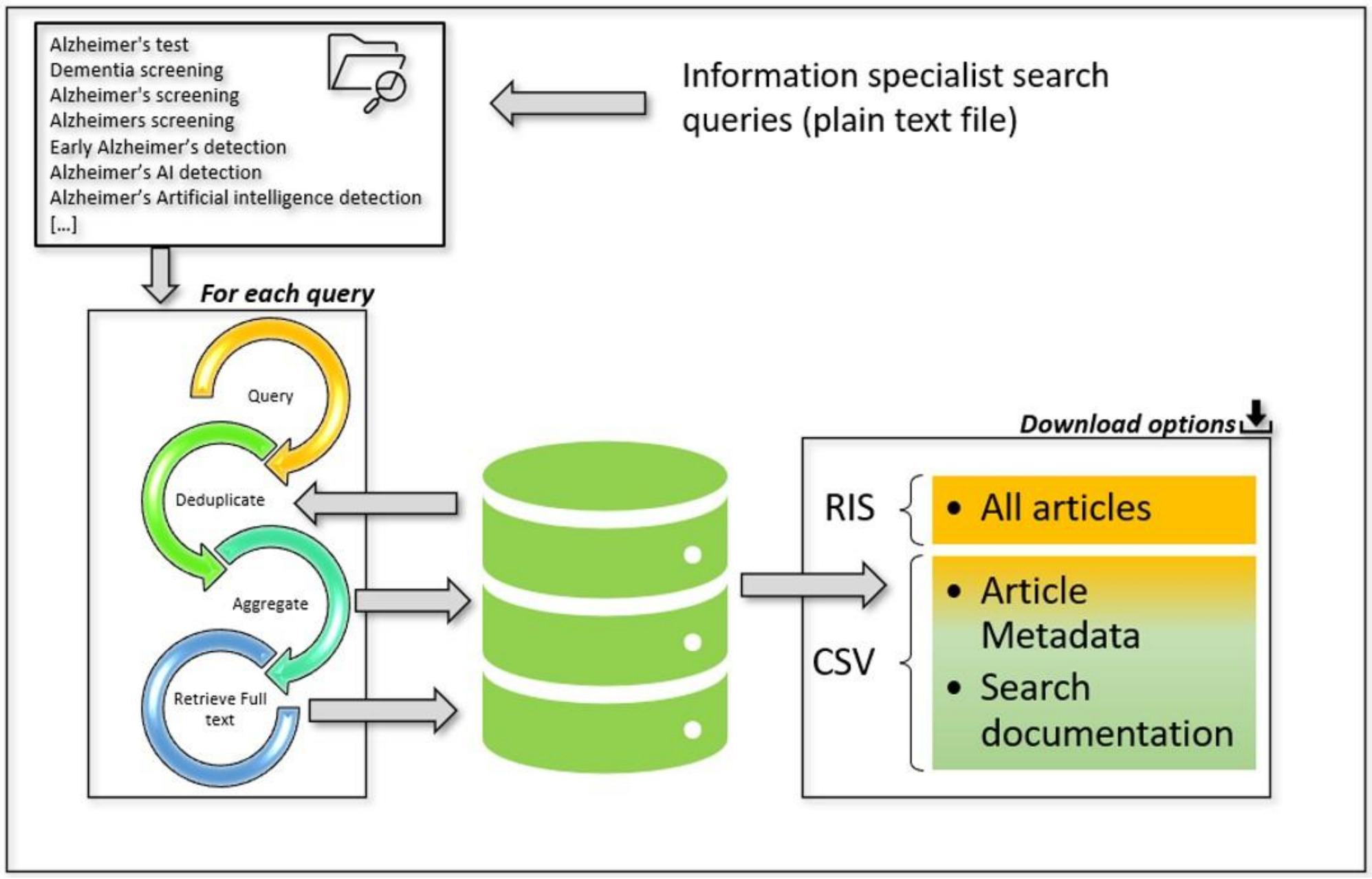

Figure 2: SCANAR flowchart showing the information retrieval process within the tool. Firstly, a text file with one query on each line is uploaded to the tool. For each query, results are retrieved, de-duplicated, aggregated, and optionally full text is retrieved. The data is stored in an internal dataframe and can be exported as RIS, CSV. Additionally, SCANAR creates an automated search documentation that shows how many results were retrieved, and how many duplicates were contained in each query.

583x372mm (47 x 47 DPI)

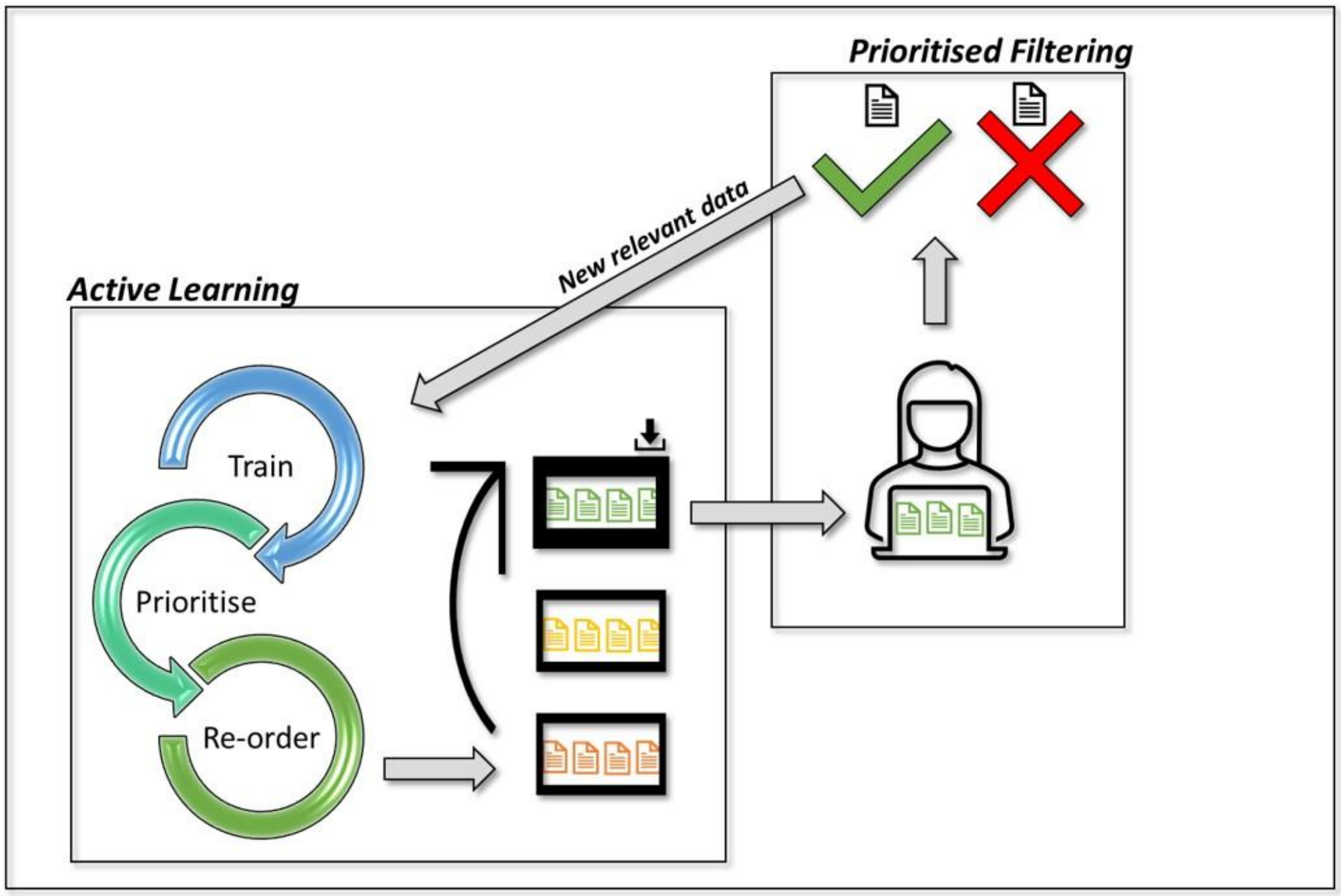

Figure 3: Flow of information through the AIDOC tool

558x372mm (47 x 47 DPI)

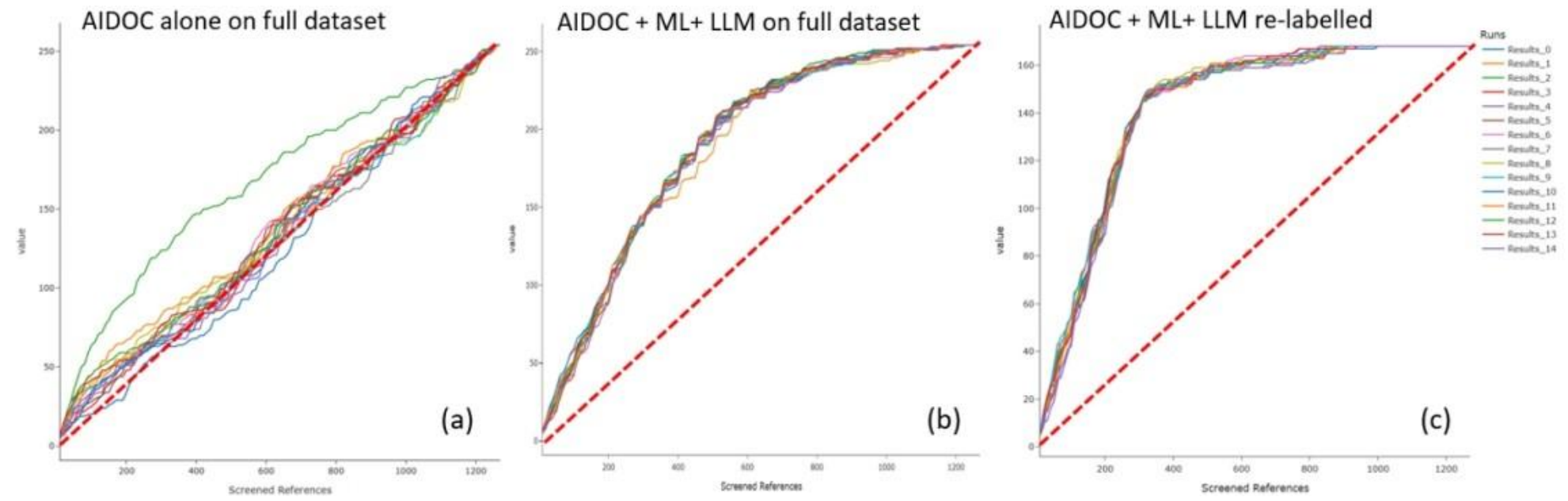

Caption : Figure 4: Performance of reference priorisation on the challenging SCANAR-derived Quantum Technology dataset, containing full text news articles across multiple horizons. The red dashed line indicates how well random sampling of articles would have performed in comparison to the AI-supported workflow. a) Shows poor performance of AIDOC alone. b) Uses LLM labels within the ensemble model to prioritise LLM-included articles. c) Re-labelled dataset, excluding vague and distant articles to visualise that they contributed most errors.

755x234mm (47 x 47 DPI)

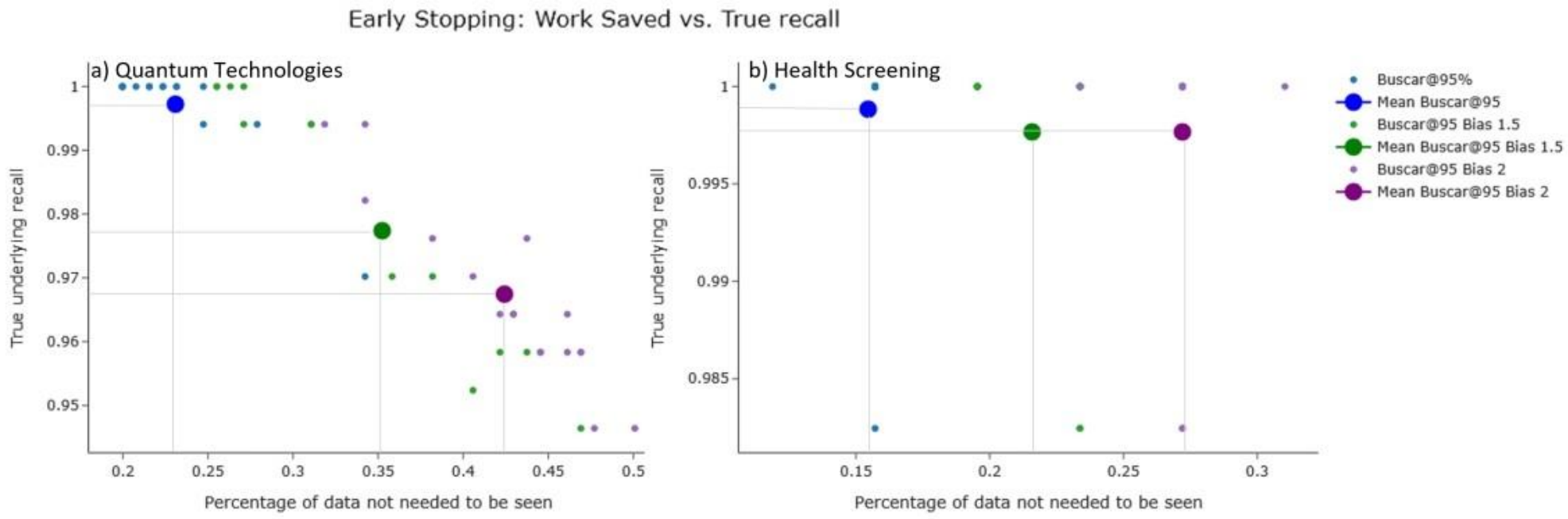

Figure 5: Results of applying CMH on the results from the iterative learning simulation study on news articles. We set it up to predict when we may safely stop screening early at 95% estimated recall, and show work savings in terms of the percentage of articles that wouldn't need to be seen on the X axis for each simulation run. They Y axis shows true underlying recall for each predicted stop of screening. Plot a) We tested three bias parameters and their impact on work-savings for the re-labelled Quantum dataset. Plot b) CMH found safe stopping thresholds with all tested parameters on the Health Screening dataset, with work-savings up to 27%.

617x234mm (130 x 130 DPI)

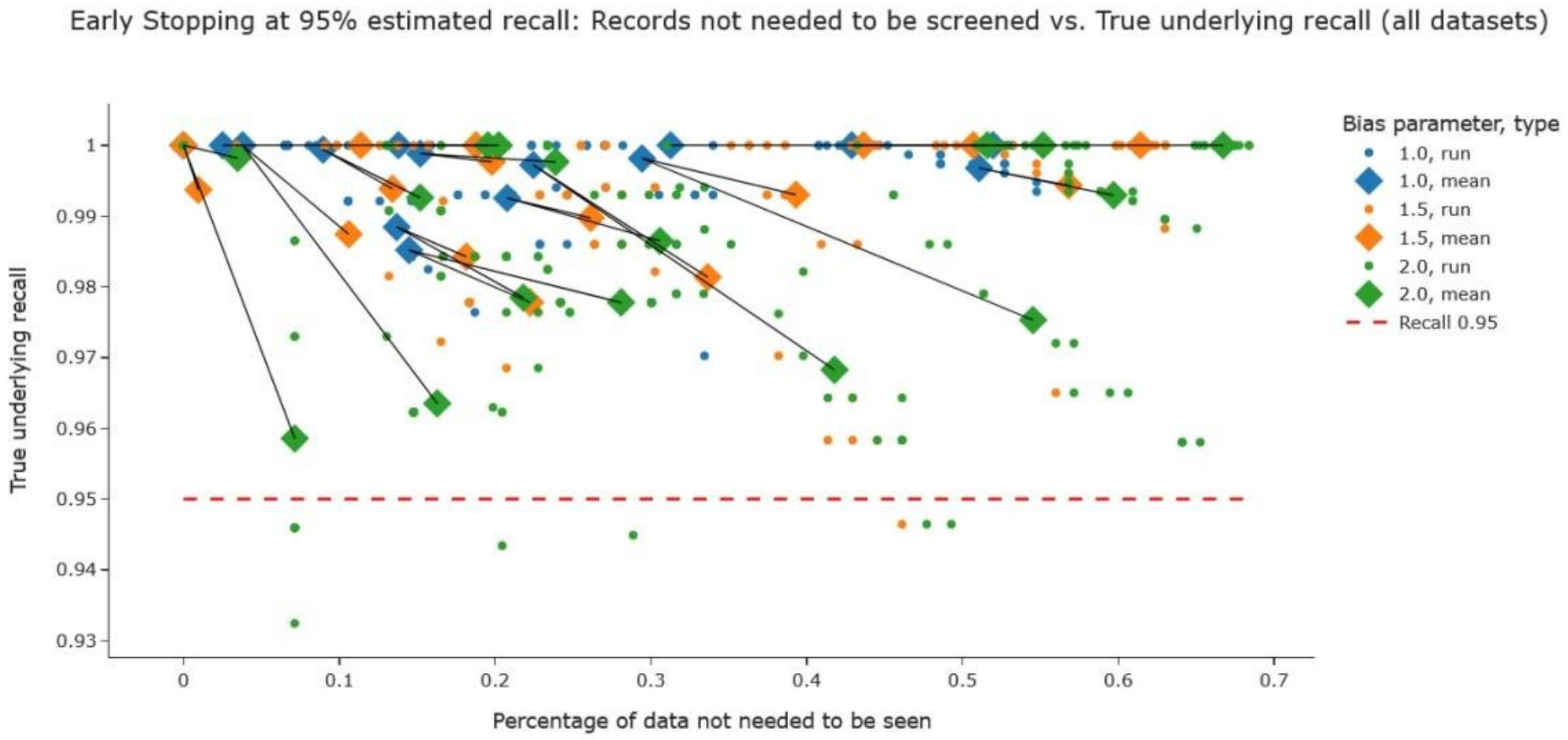

Figure 6 Results of applying CMH on the LLM ensemble classifier results. The figure displays results from all datasets across three bias parameters. Diamonds indicate the average result for a dataset, while small circles indicate exact results from runs. Average results for each dataset are connected by black lines, to show impact of changing bias parameters. The red line indicates the target recall threshold, and circles underneath this line represent instances where the algorithm predicted the stopping point too early.

714x339mm (47 x 47 DPI)

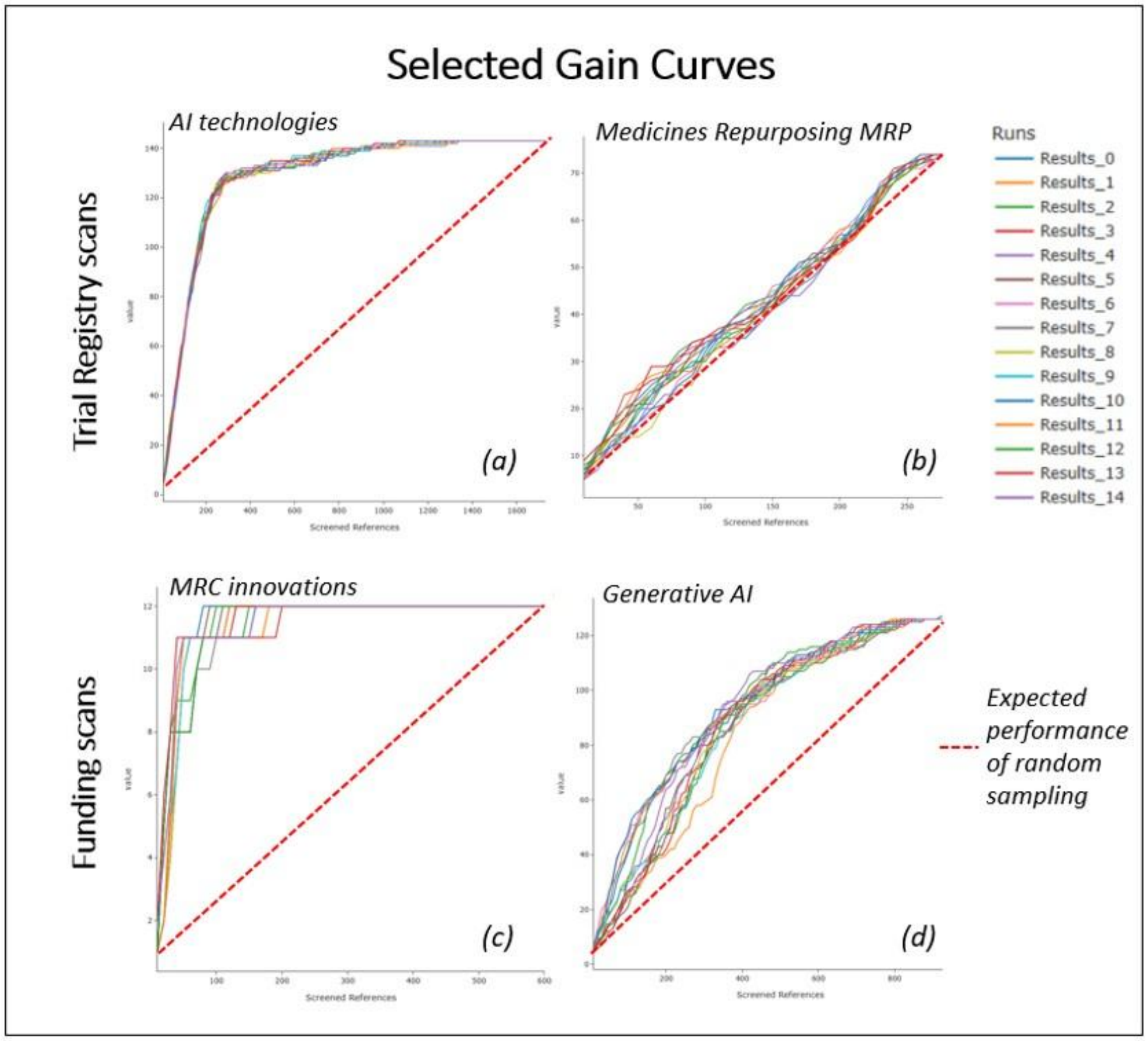

Figure 7: Gain curves, showing screening simulations across 15 random runs for 5 exemplary datasets, with comparison to the screening progress expected for random reference sampling. (a) Clinical trial registry entries about AI innovations on a dataset with good performance; (b) Clinical trial registry entries about rare and repurposed medicines showing no better than random sampling performance due to insufficient information within dataset; (c) Funding scan with MRC records showing good performance; (d) Funding scan on generative AI innovations showing below average but better than random performance.

451x409mm (47 x 47 DPI)

| Prompt Part | Content |
|---|---|
| **1: Setting the scene** | Description of research context |
| **2: Summary of relevant information** | Description of project and relevant information within references |
| **3: Exclusion criteria (optional)** | Description of irrelevant references |
| **4: Output structure** | Description of binary output format |
| **5: Context** | The relevant reference text |

**Example**

“You are a researcher screening news articles for inclusion in a literature analysis.”

“The inclusion criteria are the following: Any article that describes a newly developed or upcoming health screening method or campaign for the early detection of diseases. Screening tests can be diagnostic; to detect cancer, dementia, HPV, or any other disease and health condition within a population. Screening methods may be offered or evaluated based on a whole population or people of selected age groups and locations. Any method, such as at-home, point of care, AI-supported, analysis of biomarkers and genes, or other methods are of interest, as long as they aim to detect diseases early.”

*[insert description of irrelevant evidence if applicable, eg.: "Exclude articles that describe already established screening programmes.", "Exclude articles that screen for disease XYZ.", "Exclude news articles that solely describe published academic papers."]*

“Answer YES if the article is relevant or unclear. Answer NO if it is not. Then reproduce the exact context from the paper that contained the information on which basis you made the decision.”

“Here is the text of the article: [*insert text*]”

| Name | Text type | Horizon | Scope | Fields | Total N | N included (%) |
|---|---|---|---|---|---|---|
| **AI Technologies** | Trial registry | Emerging | Broad | Titles | 1727 | 143 (8%) |
| **MRP** | Trial registry | Imminent | Focussed | Titles | 280 | 74 (26%) |
| **GenAI Technologies 1 Trials** | Trial registry* | Transitional | Broad | Titles, Abstracts | 1141 | 143 (12%) |
| **GenAI Technologies 2 UKRI** | Funding calls* | Transitional | Broad | Titles, Abstracts | 984 | 127 (13%) |
| **GenAI Technologies 3 NIHR** | Funding calls | Transitional | Broad | Titles, Abstracts | 116 | 37 (32%) |
| **Woman Health** | Funding calls* | Emerging, Transitional | Focussed | Titles, Abstracts | 677 | 5 (0.7%) |
| **MRC funding scan** | Funding calls* | All | Broad | Titles, Abstract s | 599 | 108 (18%) |
| **Rapid Biomed 1 Genetic Engineering** | Journal Articles | Transitional, Imminent | Focussed | Titles, Abstract s | 343 | 45 (13%) |
| **Rapid Biomed 2 TX delivery** | Journal Articles | Transitional, Imminent | Focussed | Titles, Abstract s | 352 | 53 (15%) |
| **Rapid Biomed 3 Tissues Devices** | Journal Articles | Transitional, Imminent | Focussed | Titles, Abstract s | 348 | 80 (23%) |
| Quantum Technologies | News* | All | Broad | Titles, truncated full text | 1262 | 254 (20%) (sensitivity analysis version: 168 includes |
| Health Screening | News* | All | Broad | Titles, truncated full text | 261 | 57 (22%) |
| OHAT SWIFT BPA | Journal Articles | Not applicable | Focussed | Titles, Abstracts | 7700 | 111 (1.4%) |
| OHAT SWIFT Fluoride | Journal Articles | Not applicable | Focussed | Titles, Abstracts | 4479 | 51 (1.1%) |
| OHAT SWIFT Transgeneration | Journal Articles | Not applicable | Broad | Titles, Abstracts | 48632 | 765 (1.6%) |
| OHAT SWIFT PFOA- | Journal Articles | Not applicable | Focussed | Titles, Abstracts | 6328 | 95 (1.5%) |

| Name | WSS@95r | TNR@95r | Recall@50% screened | Recall@75% screened | Last included record found at |
|---|---|---|---|---|---|
| **AI Technologies** | 0.56 ± 0.03 | 0.66 ± 0.03 | 0.97 ± 0 | 1 ± 0 | 70% ± 5% |
| **MRP*** | 0.07 ± 0.02 | 0.15 ± 0.02 | 0.55 ± 0.02 | 0.78 ± 0.02 | 95 ± 2 |
| **GenAI Technologies 1 Trials** | 0.34 ± 0.04 | 0.44 ± 0.05 | 0.86 ± 0.05 | 0.99 ± 0.01 | 80 ± 4 |
| **GenAI Technologies 2 UKRI** | 0.22 ± 0.03 | 0.31 ± 0.03 | 0.83 ± 0.02 | 0.96 ± 0.01 | 97 ± 2 |
| **GenAI Technologies 3 NIHR** | 0.22 ± 0.04 | 0.39 ± 0.06 | 0.76 ± 0.06 | 0.96 ± 0.03 | 81 ± 5 |
| **Woman Health** | 0.56 ± 0.06 | 0.61 ± 0.06 | 1 ± 0 | 1 ± 0 | 39 ± 6 |
| **MRC funding scan** | 0.23 ± 0.03 | 0.33 ± 0.04 | 0.80 ± 0.02 | 96 ± 0.01 | 84 ± 3 |
| **Rapid Biomed 1 Genetic Engineering** | 0.50 ± 0.03 | 0.63 ± 0.03 | 0.97 ± 0.01 | 1 ± 0 | 53 ± 2 |
| **Rapid Biomed 2 TX delivery** | 0.35 ± 0.02 | 0.47 ± 0.02 | 0.82 ± 0.03 | 1 ± 0.0001 | 67 ± 4 |
| **Rapid Biomed 3 Tissues Devices** | 0.33 ± 0.01 | 0.48 ± 0.01 | 0.77 ± 0.02 | 1 ± 0 | 71 ± 3 |
| Quantum Technologies | 0.003± 0.01 | 0.05 ± 0.01 | 0.53 ± 0.05 | 0.76 ± 0.03 | 99 ± 0.2 |
| Health Screening | 0.35 ± 0.05 | 0.50 ± 0.07 | 0.88 ± 0.03 | 0.98 ± 0.01 | 91 ± 4 |
| OHAT SWIFT BPA | 0.73 ± 0.01 | 0.79 ± 0.01 | 1 ± 0 | 1 ± 0 | 37 ± 1 |
| OHAT SWIFT Fluoride | 0.89 ± 0.01 | 0.95 ± 0.01 | 1 ± 0 | 1 ± 0 | 24 ± 9 |
| OHAT SWIFT Transgenerational | 0.68 ± 0.01 | 0.74 ± 0.01 | 1 ± 0 | 1 ± 0 | 69 ± 3 |
| OHAT SWIFT PFOA-PFOS | 0.80 ± 0.01 | 0.86 ± 0.01 | 1 ± 0 | 1 ± 0 | 27± 5 |
| Average ± SD | 0.42 ± 0.26 | 0.51 ± 0.25 | 0.85 ± 0.15 | 0.96 ± 0.08 | 67.60 ± 24.88 |

| | OHAT SWIFT BPA | OHAT SWIFT SWIFT Fluoride | SWIFT Transgenerational PFOA- | OHAT PFOS | Notes |
|---|---|---|---|---|---|
| **Howard et al., 2016** | **0.80 ± 0.01** | 0.86 ± 0.01 | **0.74 ± 0.003** | 0.83 ± 0.01 | Using 15 seed references |
| **AIDOC** | 0.73 ± 0.01 | 0.89 ± 0.01 | 0.58 ± 0.001 | 0.70 ± 0.01 | Using 5 seed references |
| **AIDOC + SGD** | 0.73 ± 0.02 | 0.85 ± 0.01 | 0.68 ± 0.01 | 0.80 ± 0.01 | Using 5 seed references |
| **AIDOC +SGD+ LLM** | 0.77± 0.01 | **0.90 ± 0.01** | **0.74 ± 0.01** | **0.84 ± 0.01** | Using 1 seed reference |

| Name | WSS@95r | TNR@95r | Recall@50% screened | Recall@75% screened | Last included record found at |
|---|---|---|---|---|---|
| **AI Technologies** | 0.76 ± 0.02 (+20%) | 0.88 ± 0.02 (+22%) | 0.99± 0.01 (+1%) | 1 ± 0.001 | 64% ± 7% (-6%) |
| **MRP*** | 0.03 ± 0.02 (-4%) | 0.1 ± 0.03 (-5%) | 0.53 ± 0.02 (+2%) | 0.76 ± 0.02 (+2%) | 97 % ± 1% (-2%) |
| **GenAI Technologies 1 Trials** | 42 ± 0.02 (+8%) | 0.55 ± 0.03 (+11%) | 0.94 ± 0.01 (+8%) | 0.99 ± 0.01 (+0%) | 84% ± 5% (+4%) |
| **GenAI Technologies 2 UKRI** | 0.29 ± 0.03 (+7%) | 0.38 ± 0.04 (+7%) | 0.84 ± 0.02 (+1%) | 0.97 ± 0.01 (+1%) | 90± 4% (-7%) |
| **GenAI Technologies 3 NIHR** | 0.26±0.05 (+4%) | 0.44 ± 0.08 (+5%) | 0.81 ± 0.02 (+5%) | 0.97 ± 0.01 (+1%) | 83 ± 5% (+2%) |
| **Woman Health** | 0.86 ± 0.01 (+30%) | 0.92±0.01 (+31%) | 1±0 (+0%) | 1±0 (+0%) | 8 ± 1% (-31%) |
| **MRC funding scan** | 0.28 ± 0.01 (+5%) | 0.39 ± 0.02 (+6%) | 0.81 ± 0.02 (+1%) | 0.98 ± 0.01 (+2%) | 87± 1% (+3%) |
| **Rapid Biomed 1 Genetic Engineering** | 0.49 ± 0.03 (-1%) | 0.61 ± 0.03 (-2%) | 0.97 ± 0.02 (+0%) | 0.98 ± 0.0 (-2%) | 83 ± 0.4% (+30%) |
| **Rapid Biomed 2 TX delivery** | 0. 22 ± 0.06 (-13%) | 0.31 ± 0.07 (-16%) | 0.76 ± 0.05 (-6%) | 0.96 ± 0.01 (-4%) | 88 ± 6% (+13%) |
| **Rapid Biomed 3 Tissues Devices** | 0.3 ± 0.01 (-3%) | 0.44 ± 0.01 (-4%) | 0.80 ± 0.02 (+3%) | 1 ± 0 | 70 ± 2% (-1%) |
| **Quantum Technologies** | 0.29 ± 0.02 (+28%) | 0.42 ± 0.02 (+41%) | 0.88 ± 0.01 (+ 35%) | 0.97 ± 0.01 (+21%) | 95 ± 1.5 (-4%) |
| **Health Screening** | 0.49 ± 0.02 (+14%) | 0.68 ± 0.03 (+18%) | 0.96 ± 0.01 (+8%) | 0.99 ± 0.01 (+1%) | 66 ± 9 (-25%) |
| **OHAT SWIFT BPA** | 0.78 ± 0.01 (+5%) | 0.84 ± 0.01 | 1±0 (+0%) | 1±0 (+0%) | 29± 2 |
| **OHAT SWIFT Fluoride** | 90 ± 0.01 (+1%) | 0.96 ± 0.01 | 1±0 (+0%) | 1±0 (+0%) | 15 ± 1 |
| **OHAT SWIFT Transgenerational** | 0.74± 0.01 (+6%) | 0.8±0.01 (+6%) | 1±0 (+0%) | 1±0 (+0%) | 63 ± 4 |
| **OHAT SWIFT PFOA-PFOS** | 0.84 ± 0.01 (+4%) | 0.91 ± 0.01 | 1±0 (+0%) | 1±0 (+0%) | 14 ± 1 |
| **Average ± SD (all)** | 0.50 ± 0.27 (+4%) | 0.60 ± 0.25 (+5%) | 0.89 ± 0.12 (+2%) | 0.97 ± 0.05 (+1%) | 64 ± 0.29 (-3%) |

| Name | WSS@95r | TNR@95r | Recall@50% screened | Recall@75% screened | Last included record found at | Note |
|---|---|---|---|---|---|---|
| Health Screening LLM + ML +AIDO | 0.49 ± 0.02 (+14%) | 0.68 ± 0.03 (+18%) | 0.96 ± 0.01 (+8%) | 0.99 ± 0.01 (+1%) | 66 ± 9 (-25%) | Using full text |
| Quantum Technology LLM+ ML + AIDOC | 0.29 ± 0.02 (+28%) | 0.42 ± 0.02 (+41%) | 0.88 ± 0.01 (+35%) | 0.97 ± 0.01 (+21%) | 95 ± 1.5 (-4%) | Using full text, full original dataset |
| Quantum Technology (imminent/transient) LLM + | 0.50 (± 0.04) (+49%) | 0.63 (± 0.05) (+62%) | 0.96 ± 0.01 (+43%) | 0.99 ± <0.01 (+24%) | 70 ± 3 (-29%) | Sensitivity/ Error analysis |